\documentclass{tlp}
\usepackage{latexsym}
\usepackage{amssymb}
\usepackage{aopmath}
\usepackage{pslatex}
\title[Inference of termination conditions]
{Inference of termination conditions for numerical loops in Prolog} 
\author[Alexander Serebrenik and Danny De Schreye]
         {ALEXANDER SEREBRENIK and DANNY DE SCHREYE\\
         Department of Computer Science, K.U. Leuven\\
Celestijnenlaan 200A, B-3001, Heverlee,
Belgium\\
         \email{\{Alexander.Serebrenik,Danny.DeSchreye\}@cs.kuleuven.ac.be}}

\newtheorem{example}{Example}
\newtheorem{definition}{Definition}

\begin{document}
\newcommand{\eat}[1]{}
\maketitle

\begin{abstract}
We present a new approach to termination analysis of numerical
computations in logic programs. Traditional approaches fail to
analyse them due to non well-foundedness of
the integers. We present a technique that allows overcoming these
difficulties. Our approach is based on transforming a program in a way 
that allows integrating and extending techniques 
originally developed for analysis of 
numerical computations in the framework of query-mapping pairs 
with the well-known framework of acceptability.
Such an integration not only contributes to the understanding of 
termination behaviour of numerical computations, but also
allows us to perform a correct analysis of such computations automatically,
by extending previous work on a constraint-based approach to termination.
Finally, we discuss possible extensions of 
the technique, including incorporating general term orderings.

\end{abstract}

\section{Introduction}
One of the important aspects in verifying the correctness of logic programs 
(as well as functional programs and term rewrite systems) is verification of
termination. Due to the declarative formulation of programs, the danger of 
non-termination may be increased. As a result, termination analysis received
considerable attention in logic programming (see e.g.~\cite{Apt:Marchiori:Palamidessi,Bossi:Cocco:Etalle:Rossi:modular,Bruynooghe:Codish:Genaim:Vanhoof,Codish:Taboch,Colussi:Marchiori:Marchiori,Decorte:DeSchreye:Vandecasteele,Dershowitz:Lindenstrauss:Sagiv:Serebrenik,Genaim:Codish:Gallagher:Lagoon,Lindenstrauss:Sagiv,Mesnard,Mesnard:Ruggieri,Plumer,Ruggieri:CLP,Verbaeten:Sagonas:DeSchreye}).

Numerical computations form an essential part of almost any 
real-world program. Clearly, in order for a termination analyser to be of
practical use it should contain a mechanism for inferring termination
of such computations. However, this topic attracted less attention of
the research community. In this paper we concentrate on automatic
termination inference for logic programs depending on numerical computations.

Dershowitz {\em et al.}~\cite{Dershowitz:Lindenstrauss:Sagiv:Serebrenik}
showed that termination of general numerical computations,
for instance on floating point numbers, may be 
counter-intuitive, i.e., the observed behaviour does not necessarily
coincide with the theoretically expected one. Moreover, as the following
program shows, similar results can be obtained even if the built-in
predicates of the underlying language are restricted to include ``greater
than'' and multiplication only.
\begin{example}
Consider the following program, that given a positive number
$x$ results in
a sequence of calls 
$p(0\mbox{.}25x), p((0\mbox{.}25)^2x), \ldots$ 
\[\begin{array}{l}
 p(X) \leftarrow X > 0, X1\;\;\mbox{\sl is}\;\;X*0\mbox{.}25, p(X1)\mbox{.}
\end{array}\]
If we reason purely
in terms of real numbers, we might expect
that the computation started by $p(1\mbox{.}0)$ will be infinite.
However, in practice the goal above terminates with respect to 
this program, since there exists $k$, such that $(0\mbox{.}25)^k$ is small
enough for the comparison $(0\mbox{.}25)^k > 0$ to fail.
$\hfill\Box$\end{example}
We discuss these issues in detail in~\cite{Serebrenik:DeSchreye:real}.
In the current paper we avoid these complications by restricting to 
integer computations only.

Next, we illustrate the termination problem for integer computations
with the following example:
\begin{example}
\label{example:loop:to7}
Consider the following program:
\[\begin{array}{l} 
 p(X) \leftarrow X < 7, X1\;\mbox{\sl is}\;X+1, p(X1)\mbox{.}
\end{array}\]
This program terminates for queries $p(X)$, for all integer values
of $X$. 
$\hfill\Box$\end{example}
Most of the existing automated approaches to termination analysis for logic
programs~\cite{Codish:Taboch,Lindenstrauss:Sagiv,Mesnard:Neumerkel,Ohlebusch} 
fail to prove termination for such examples. The reason is that they are most often based on the notion of a {\em level mapping}, that is, a function from the set of all possible atoms to the natural numbers, which should decrease while traversing the rules. Usually level mappings are defined to depend on the structure of terms and to ignore constants, making the analysis of Example~\ref{example:loop:to7} impossible. 

Of course, this can be easily repaired, by considering level mappings that map each natural number to itself. In fact, the kernels of two termination analysers for logic programs, namely cTI~\cite{Mesnard,Mesnard:Neumerkel} and TerminWeb~\cite{Codish:Taboch}, rely on abstracting logic programs to CLP(${\cal N}$) programs, and use the identity level mapping on ${\cal N}$ in the analysis of the abstract versions of the programs\footnote{We thank anonymous referees for pointing this link to related work out to us.}.

Note however that this is insufficient for the analysis of Example~\ref{example:loop:to7}. In fact, there remain two problems. First, the program in Example~\ref{example:loop:to7} is defined on a (potentially negative) integer argument. This means that we need a level mapping which is different from the identity function.

Two approaches for solving this problem are possible. First, one can
change the definition of the level mapping to map atoms to integers. However,
integers are not well-founded. To prove termination
one should prove that the mapping is to some well-founded subset of integers.
In the example above $(-\infty,7)$ forms such a subset with an ordering
$\succ$, such that $x \succ y$ if $x < y$, with respect to the usual ordering
on integers. Continuing this line of thought one might consider mapping
atoms to more general well-founded domains. In fact, already in the early days
of program analysis~\cite{Floyd,Katz:Manna} general well-founded domains were discussed. However, the growing importance of automatic termination analysers
and requirements of robustness and efficiency stimulated researchers to look 
for more specific instances of well-founded domains, such as natural numbers
in logic programming and terms in term-rewriting systems.

The second approach, that we present in the paper, does not require
changing the definition of level mapping. Indeed,
the level mapping as required exists. It maps
$p(X)$ to $7-X$ if $X<7$ and to $0$ otherwise. This level mapping decreases
while traversing the rule, i.e., the size of $p(X)$, $7-X$ for $X<7$,
is greater than the size of $p(X1)$, $6-X$ for $X < 7$ and $0$ for $X\geq 7$, 
thus, proving termination.

A second problem with approaches based on the identity function, as the level 
mapping used on CLP(${\cal N}$), is that, even if the program in Example~\ref{example:loop:to7} would have been defined on natural values of $X$ only, they would still not be able to prove termination. The reason is that the natural argument increases under the standard ordering of the natural numbers. Such bounded increases (be it of structure-sizes or of numerical values) are not dealt with by standard termination analysers. Note that the two approaches presented above also solve this second problem.

The main contribution of this paper is that we provide a transformation - similar to multiple specialisation~\cite{Winsborough} - that allows us to define level mappings of the form illustrated in the second approach above in an automatic way. To do so, we incorporate techniques of~\cite{Dershowitz:Lindenstrauss:Sagiv:Serebrenik}, such as
level mapping inference, in the framework of the acceptability
with respect to a set~\cite{DeSchreye:Verschaetse:Bruynooghe,%
Decorte:DeSchreye:98}. This integration provides not only a 
better understanding of termination behaviour of integer computations, 
but also the possibility to perform the analysis automatically
as in Decorte {\em et al.}~\cite{Decorte:DeSchreye:Vandecasteele}.

Moreover, we will also be somewhat more general 
than~\cite{Decorte:DeSchreye:Vandecasteele}, by studying the problem of
termination inference, rather than termination verification. More
precisely, we will be inferring conditions that, if imposed on the
queries, will ensure that the queries will terminate. Inference of 
termination conditions was studied in~\cite{Mesnard,Mesnard:Neumerkel,Genaim:Codish}. Unlike termination conditions inferred by these
approaches, stated in terms of groundedness of arguments, our technique 
produces conditions based on domains of the arguments, as shown in
Example~\ref{example:loop:to7b}. 

\begin{example}
\label{example:loop:to7b}
Extend the program of Example~\ref{example:loop:to7} with the following
clause:
\[\begin{array}{l}
 p(X) \leftarrow X > 7, X1\;\mbox{\sl is}\;X+1, p(X1)\mbox{.}
\end{array}\]
This extended program terminates for $X\leq 7$ and this is the 
condition we will infer.
$\hfill\Box$\end{example}

The rest of the paper is organised as follows. After making some
preliminary remarks, we present in Section 3 our transformation---first 
by means of an example, then more formally. In Section 4 we discuss 
more practical issues and present  the algorithm implementing the
termination inference. Section 5 contains the results of an experimental
evaluation of the method.
In Section 6 we discuss further extensions,
such as proving termination of programs depending on numerical computations
as well as symbolic ones.
We summarise our contribution in Section 7, review related work and conclude.

\section{Preliminaries}
We follow the standard notation for terms and atoms. A {\em query} is a 
finite sequence of atoms. Given an atom $A$, $\mbox{\sl rel}(A)$ denotes
the predicate occurring in $A$. $Atom_P$ ($Term_P$)
denotes the set of all atoms (terms) that can be constructed from
the language underlying $P$. The extended Herbrand Base $B^E_P$ 
(the extended Herbrand Universe $U^E_P$) 
is the quotient set of $Atom_P$ ($Term_P$) modulo the variant relation. 
An SLD-tree constructed using the left-to-right selection rule of
Prolog is called an LD-tree. A goal $G$ {\em LD-terminates} for
a program $P$, if the LD-tree for $(P,G)$ is finite. 

The following definition is 
similar to Definition 6.30~\cite{Apt:Book}.
\begin{definition}
Let $P$ be a program and $p$, $q$ be predicates occurring in it. We say that
\begin{itemize}
\item {\it $p$ refers to $q$ in $P$\/} if there is a clause in $P$
that uses $p$ in its head and $q$ in its body.
\item {\it $p$ depends on $q$ in $P$\/} and write $p\sqsupseteq q$,
if $(p,q)$ is in the transitive 
closure of the relation {\it refers to}.
\item {\it $p$ and $q$  are mutually recursive\/} and write $p\simeq q$, if $p\sqsupseteq q$ and $q\sqsupseteq p$.
\end{itemize} 
\end{definition}
The only difference between our definition and the one by Apt~\cite{Apt:Book}
is that we require the relation $\sqsupseteq$ to be the {\em transitive\/}
closure of the relation {\em refers to}, while~\cite{Apt:Book} requires
it to be {\em transitive, reflexive\/} closure. Using our definition
we call a predicate $p$ {\em recursive\/} if $p\simeq p$ holds.

We recall some basic notions, related to termination
analysis. A {\em level mapping} is a function 
$\mid\cdot\mid: B^E_P\rightarrow {\cal N}$, where ${\cal N}$ is the set of 
the naturals. Similarly, a {\em norm} is a function 
$\|\cdot\|: U^E_P\rightarrow {\cal N}$.

We study termination of programs with respect to sets of
queries. The following notion is one of the most basic notions 
in this framework.
\begin{definition}
Let $P$ be a definite program and $S$ be a set of atomic queries. 
The {\em call set}, $\mbox{\sl Call}(P,S)$, is the set of all atoms $A$
from the extended Herbrand Base $B^E_P$,
such that a variant of $A$ is a selected atom in some derivation for
$P\cup \{\leftarrow Q\}$, for some $Q\in S$ and under the left-to-right 
selection rule.
\end{definition}

The following definition~\cite{Serebrenik:DeSchreye:LOPSTR2000} generalises the notion of acceptability with 
respect to a set~\cite{DeSchreye:Verschaetse:Bruynooghe,Decorte:DeSchreye:98} 
by extending it to mutual recursion.
\begin{definition}
Let $S$ be a set of atomic queries and $P$ a definite  
program. $P$ is {\em acceptable with respect to $S$} if there exists a 
level mapping $\mid\cdot\mid$ such that 
\begin{itemize}
\item for any $A\in\mbox{\sl Call}(P,S)$
\item for any clause $A'\leftarrow B_1,\ldots,B_n$ in $P$, such that 
$\mbox{\rm mgu}(A,A') = \theta$ exists,
\item for any atom 
$B_i$, such that $\mbox{\sl rel}(B_i)\simeq \mbox{\sl rel}(A)$
and for any computed answer substitution $\sigma$ for 
$\leftarrow (B_1, \ldots, B_{i-1})\theta$ holds that
\[\mid A \mid\;>\;\mid B_i\theta\sigma \mid.\]
\end{itemize}
\end{definition}
De Schreye {\em et al.}~\cite{DeSchreye:Verschaetse:Bruynooghe} characterise
LD-termination in terms of acceptability.
\begin{theorem}[cf.\ ~\cite{DeSchreye:Verschaetse:Bruynooghe}]
\label{acc:term}
Let $P$ be a definite program. $P$ is acceptable with respect to a set 
of atomic queries $S$ if and only if $P$ is LD-terminating for all queries in $S$.
\end{theorem}

We also need to introduce notions of rigidity and of interargument relations.
Given a norm $\|\cdot\|$ and a term $t$, Bossi {\em et al.}~\cite{Bossi:Cocco:Fabris:TAPSOFT}
call $t$ {\em rigid} with respect to $\|\cdot\|$ if for any substitution $\sigma$, 
$\|t\sigma\| = \|t\|$. Observe that ground terms are rigid with respect to all norms. The notion of rigidity is obviously extensible to atoms and level mappings. Interargument relations have initially been studied by~\cite{Ullman:van:Gelder,Plumer,Verschaetse:DeSchreye}. In this paper we use the definition of~\cite{Decorte:DeSchreye:Vandecasteele}.

\begin{definition}
Let $P$ be a definite program, $p/n$ a predicate in $P$. 
An {\em interargument relation} for $p/n$ is a relation 
$R_p \subseteq {\cal N}^n$. 
$R_p$ is a {\em valid interargument relation for $p/n$ with respect to a norm
$\|\cdot\|$} if and only if for every 
$p(t_1,\ldots,t_n)\in \mbox{\sl Atom}_P$ if 
$P\models p(t_1,\ldots,t_n)$ then $(\|t_1\|,\ldots,\|t_n\|)\in R_p$. 
\end{definition}

Combining the notions of rigidity, acceptability and interargument relations
allows us to reason on termination completely at the clause level. 
\begin{theorem}[rigid acceptability (cf.\ ~\cite{Decorte:DeSchreye:Vandecasteele})]
\label{rigid:acc:term}
Let $S$ be a set of atomic queries and $P$ a definite program. Let $\|\cdot\|$
be a norm and, for each predicate $p$ in $P$, let $R_p$ be a valid interargument relation
for $p$ with respect to $\|\cdot\|$. If there exists a level mapping $\mid\cdot\mid$ which is
rigid on $\mbox{\sl Call}(P,S)$ such that
\begin{itemize}
\item for any clause $H\leftarrow B_1,\ldots,B_n\in P$, and
\item for any atom $B_i$ in its body such that $\mbox{\sl rel}(B_i)\simeq \mbox{\sl rel}(H)$,
\item for substitution $\theta$ such that the arguments of the atoms in 
$(B_1,\ldots,B_{i-1})\theta$ all satisfy their associated interargument relations $R_{B_1},
\ldots, R_{B_{i-1}}$:
\[\mid H\theta\mid > \mid B_i\theta\mid \]
then $P$ is acceptable with respect to $S$.
\end{itemize}
\end{theorem}

\eat{
To characterise program transformations Bossi and Cocco~\cite{Bossi:Cocco}
introduced the following notion for a program $P$ and a query $Q$: 
%
\begin{eqnarray*}
 {\cal M}[\![P]\!](Q) = \\
 \hspace{1.0cm} \{\sigma\mid\mbox{\rm there is a successful LD-derivation of $Q$ and $P$ with c.a.s. $\sigma$}\} \\
&& \hspace{1.0cm} \cup \{\bot\mid\mbox{\rm there is an infinite LD-derivation of $Q$ and $P$}
\} 
\end{array}\]
} 

\section{Methodology}
\label{section:methodology}
In this section we introduce our methodology using a simple example.
In the subsequent sections, we formalise it and discuss 
different extensions.

Computing a query with respect to the following example results in 
a sequence of calls with oscillating arguments like $p(-2), p(4), p(-16), \ldots$
and stops if the argument is greater than $1000$ or smaller than 
$-1000$. The treatment is done first on the intuitive level.
\begin{example}
\label{example:osc}
We are interested in proving termination of the set of queries 
$\{p(z)\mid z\;\mbox{\rm is an integer}\}$ 
with respect to the following program:
\[\begin{array}{l}
 p(X) \leftarrow X > 1, X < 1000, X1\;\mbox{\sl is}\;-X*X, p(X1)\mbox{.} \\
 p(X) \leftarrow X < -1, X > -1000, X1\;\mbox{\sl is}\;X*X, p(X1)\mbox{.} 
\end{array}\]
The direct attempt to define the level mapping of $p(X)$ as $X$ fails,
since $X$ can be positive as well as negative. Thus, a more complex
level mapping should be defined. We start with some observations.

The first clause is applicable if $1 < X < 1000$, the second one,
if $-1000 < X < -1$. Thus, termination of $p(X)$ for
$X\leq -1000$, $-1 \leq X \leq 1$ or $X\geq 1000$ is trivial.
Moreover, if the first clause is applied and $1 < X < 1000$ holds, then 
either $-1000 < X1 < -1$ or $X1 \leq -1000 \vee -1 \leq X1 \leq 1 \vee X1 \geq 1000$ 
should hold. Similarly, if the second clause is applied and $-1000 < X < 1$ 
holds, either $1 < X1 < 1000$ or $X1 \leq -1000 \vee -1 \leq X1 \leq 1 \vee X1 \geq 1000$
should hold. 

We use this observation and split the domain of the argument of $p$,
denoted $p_1$, in three parts as following:
\[
\begin{array}{ll}
\mbox{\sf a} & 1 < p_1 < 1000\\
\mbox{\sf b} & -1000 < p_1 < -1\\
\mbox{\sf c} & p_1 \leq -1000 \vee -1 \leq p_1 \leq 1 \vee p_1 \geq 1000
\end{array}
\]
Next we
replace the predicate $p$
with three new predicates $p^{\mbox{\sf a}}$, $p^{\mbox{\sf b}}$ and
$p^{\mbox{\sf c}}$. We add conditions before the 
calls to $p$ to ensure that 
$p^{\mbox{\sf a}}$ is called if $p(X)$ is called and $1 < X < 1000$
holds, $p^{\mbox{\sf b}}$ is called if $p(X)$ is called and 
$-1000 < X < -1$ holds and $p^{\mbox{\sf c}}$
is called if $p(X)$ is called and $X \leq -1000 \vee -1 \leq X \leq 1 \vee X \geq 1000$
holds. The following program is obtained:
\[\begin{array}{l}
 p^{\mbox{\sf a}}(X) \leftarrow X > 1, X < 1000, X1\;\mbox{\sl is}\;-X*X, \\
 \hspace{1.0cm} -1000 < X1, X1 < -1, p^{\mbox{\sf b}}(X1)\mbox{.} \\
 p^{\mbox{\sf a}}(X) \leftarrow X > 1, X < 1000, X1\;\mbox{\sl is}\;-X*X, \\
 \hspace{1.0cm} (X1 \leq -1000 ; (-1 \leq X1, X1 \leq 1) ; X1\geq 1000), 
 p^{\mbox{\sf c}}(X1)\mbox{.} \\
 p^{\mbox{\sf b}}(X) \leftarrow X < -1, X > -1000, X1\;\mbox{\sl is}\;X*X, \\
 \hspace{1.0cm} 1 < X1, X1 < 1000, p^{\mbox{\sf a}}(X1)\mbox{.} \\
 p^{\mbox{\sf b}}(X) \leftarrow X < -1, X > -1000, X1\;\mbox{\sl is}\;X*X, \\
 \hspace{1.0cm} (X1 \leq -1000 ; (-1 \leq X1, X1 \leq 1) ; X1\geq 1000), 
 p^{\mbox{\sf c}}(X1)\mbox{.}
\end{array}\]
Observe that the transformation we performed is a form of 
multiple specialisation, well-known in the context of abstract 
interpretation~\cite{Winsborough}.

Now we define three {\em different} level mappings, one for atoms of
$p^{\mbox{\sf a}}$, another one for atoms of $p^{\mbox{\sf b}}$
and the last one for atoms of $p^{\mbox{\sf c}}$. 
Let 
\[
\begin{array}{lcl}
\mid p^{\mbox{\sf a}}(n)\mid &=& 
\left\{
\begin{array}{ll}
1000 - n & \mbox{\rm if $1 < n < 1000$}\\
0        & \mbox{\rm otherwise}
\end{array}
\right. \\
\mid p^{\mbox{\sf b}}(n)\mid &=& 
\left\{
\begin{array}{ll}
1000 + n & \mbox{\rm if $-1000 < n < -1$}\\
0        & \mbox{\rm otherwise}
\end{array}
\right. \\
\mid p^{\mbox{\sf c}}(n)\mid &=& 0
\end{array}
\]
We verify acceptability of the 
transformed program  with respect to 
$\{p^{\mbox{\sf a}}(n) \mid 1 < n < 1000\} \cup
\{p^{\mbox{\sf b}}(n)\mid -1000 < n < -1\}$ via the 
specified level mappings.
This implies termination of the transformed program with respect to
these queries, and thus, termination of the original program with respect to 
$\{p(z)\mid z\;\mbox{\rm is an integer}\}$.

For the sake of brevity we discuss only 
queries of the form $p^{\mbox{\sf a}}(n)$ for 
$1 < n < 1000$. Heads of the first and the second clauses 
can be unified with this query, however, the second clause
does not contain calls to predicates mutually recursive with 
$p^{\mbox{\sf a}}$ and the only such atom in the first clause
is $p^{\mbox{\sf b}}(m)$, where $m = -n^2$. Then, 
$\mid p^{\mbox{\sf a}}(n)\mid\;>\;\mid p^{\mbox{\sf b}}(m)\mid$
should hold, i.e.,
$1000 - n > 1000 + m$, that is $1000 - n > 1000 - n^2$ 
($n > 1$ and $m = -n^2$), which is true for $n > 1$. 
\eat{
. Since $n > 1$ and $m = -n^2$, we require that
$1000 - n > 1000 - n^2$ which is true for $n > 1$. 
} 

For queries of the form 
$p^{\mbox{\sf b}}(n)$, the acceptability
condition is reduced to $1000 + n > 1000 - n^2$ which is true for $n < -1$.
$\hfill\Box$\end{example}

The intuitive presentation above hints at the main issues to be discussed
in the following sections: how the cases such as those 
above can be extracted from the program, and how given the extracted cases, 
the program should be transformed.
Before discussing the answers to these questions we present some basic
notions.

\subsection{Basic notions}

In this section we formally introduce some notions 
that further analysis will be based on.
Recall that the aim of our analysis is to find, given a predicate and a query,
a sufficient condition for termination of this query with respect to
this program. Thus, we need to define a notion of a termination condition.
We start with a number of auxiliary definitions.

Given a predicate $p$, $p_i$ denotes the $i$-th argument of $p$ and is
called {\em argument position denominator}. 
\begin{definition}
Let $P$ be a program, $S$ be a set of queries. An argument position $i$
of a predicate $p$ is called {\em integer argument position}, if for 
every $p(t_1,\ldots,t_n)\in \mbox{\sl Call}(P,S)$, $t_i$ is an integer.
\end{definition}
Argument position denominators corresponding to integer argument positions
will be called {\em integer argument position denominators}.

An {\sl integer inequality} is an atom of one of the following forms
$\mbox{\sl Exp1} > \mbox{\sl Exp2}$,
$\mbox{\sl Exp1} < \mbox{\sl Exp2}$,$\mbox{\sl Exp1} \geq \mbox{\sl Exp2}$ or
$\mbox{\sl Exp1} \leq \mbox{\sl Exp2}$, where $\mbox{\sl Exp1}$ and
$\mbox{\sl Exp2}$ are constructed from
integers, variables and the four operations of arithmetics on integers. A {\sl symbolic
inequality over the arguments of a predicate $p$} is constructed similarly 
to an integer inequality. However, instead of variables, integer 
argument positions denominators are used. 

\begin{example}
$X > 0$ and $Y \leq X + 5$ are integer inequalities. Given a predicate $p$
of arity 3, having only integer argument positions, 
$p_1 > 0$ and $p_2 \leq p_1 + p_3$ are symbolic 
inequalities over the arguments of $p$.
$\hfill\Box$\end{example}

Disjunctions of conjunctions based on integer inequalities are called 
{\sl integer conditions}. Similarly,  disjunctions of conjunctions
based on symbolic inequalities over the arguments of the same predicate 
are called {\sl symbolic conditions over the 
integer arguments of this predicate}. 
\eat{
\begin{example}
$X > 0 \wedge Y \leq X + 5$ is an integer condition. Given a predicate $p$
as above $p_1 > 0 \wedge p_2 \leq p_1 + p_3$ is a symbolic condition
over the integer arguments of $p$.
$\hfill\Box$\end{example}
}
\begin{definition}
Let $p(t_1,\ldots,t_n)$ be an atom and let $c_p$ be a symbolic condition
over the arguments of $p$. An {\em instance of the condition with respect to an atom}, denoted 
$c_p(p(t_1,\ldots,t_n))$,
is obtained by replacing the 
argument positions denominators with the corresponding
arguments, i.e., $p_i$ with $t_i$.
\end{definition}

\begin{example}
Let $p(X, Y, 5)$ be an atom and let $c_p$ be a symbolic condition
$(p_1 > 0) \wedge (p_2 \leq p_1 + p_3)$. Then, $c_p(p(X, Y, 5))$
is $(X > 0) \wedge (Y \leq X + 5)$.
$\hfill\Box$\end{example}

Now we are ready to define {\em termination condition\/} formally.

\begin{definition}
Let $P$ be a program, and $Q$ be an atomic query.
A symbolic condition $c_{\mbox{\sl rel}(Q)}$ is a 
{\em termination condition} for $Q$ if given that $c_{\mbox{\sl rel}(Q)}(Q)$
holds, $Q$ left-terminates with respect to $P$.
\end{definition}

For any integer $z$ a termination condition for $p(z)$ with respect to 
Examples~\ref{example:loop:to7} 
and~\ref{example:osc} is {\sl true}, i.e., for any integer $z$, 
$p(z)$ terminates with respect to these programs. 
Clearly, more than one termination condition is possible for a given query 
with respect to a given program. For example, termination conditions for 
$p(5)$ with respect to Example~\ref{example:loop:to7b}, are
among others, {\sl true}, $p_1\leq 7$ and $p_1 > 0$. Analogously,
{\sl false}, $p_1\leq 7$, $p_1 < 10$ are termination conditions for 
$p(11)$ with respect to Example~\ref{example:loop:to7b}. It should
also be noted that a disjunction of two termination conditions is always
a termination condition.

Similarly to Theorems~\ref{acc:term} and ~\ref{rigid:acc:term}
we would like to consider termination with respect to sets of atomic
queries. Therefore we extend the notion of termination condition to
a set of queries. This, however, is meaningful only if all the queries
of the set have the same predicate. We call such a set {\em single predicate
set of atomic queries}. For a single predicate
set of atomic queries $S$, $\mbox{\sl rel}(S)$ denotes the predicate
of the queries of the set.

\begin{definition}
Let $P$ be a program, and $S$ be a single predicate set of atomic queries.
A symbolic condition $c_{\mbox{\sl rel}(S)}$ is a 
{\em termination condition} for $S$ if $c_{\mbox{\sl rel}(S)}$ is a 
termination condition for all $Q\in S$.
\end{definition}

From the discussion above it follows that a termination condition for
$S = \{p(z)\mid z\;\mbox{\rm is an}$ $\mbox{\rm integer}\}$ with respect to 
Examples~\ref{example:loop:to7} and~\ref{example:osc} is {\sl true}.
This is not the case for Example~\ref{example:loop:to7b}, since
termination is observed only for some queries of $S$, namely
$p(z)$, such that $z\leq 7$. Thus, $p_1\leq 7$ is a termination
condition for $S$ with respect to Example~\ref{example:loop:to7b}.

\eat{
\begin{example}
\label{example:simple}
$q(X) \leftarrow X > 0, X \leq 5, q(X)\mbox{.}\;\;q(X) \leftarrow X > -5\mbox{.}$
This program terminates for $q(n)$, such that $n \leq -5$,
since no rule is applicable, or $-5 < n \leq 0 \vee n > 5$, since 
repeated rule application
in this case is finite. Thus, a termination condition for the goal
$\leftarrow q(X)$ is $q_1\leq 0 \vee q_1 > 5$.
$\hfill\Box$\end{example}
} 

We discuss now inferring what values  integer arguments can take
during traversal of the rules, i.e., the ``case analysis'' performed
in Example~\ref{example:osc}. 
It provides already 
the underlying 
intuition---calls of the predicate
$p^c$ are identical to the calls of the predicate $p$, where $c$ holds
for its arguments. More formally, we define a notion 
of a {\em set of adornments}.
Later  we specify when it is {\em guard-tuned} and we show how such a 
guard-tuned set of adornments can be constructed.
\eat{
\begin{definition}
\label{def:set:of:adornments}
Let $p$ be a predicate, and let $c_1,\ldots,c_n$ be symbolic conditions
over the integer arguments of $p$. The set ${\cal A}_p = \{c_1,\ldots,c_n\}$
is called {\em set of adornments for $p$} if $\bigvee_{i=1}^n c_i = \mbox{\sl true}$
and for all $i,j$ such that 
$1\leq i < j\leq n$, $c_i \wedge c_j = \mbox{\sl false}$.
\end{definition}
}
\begin{definition}
\label{def:set:of:adornments}
Let $p$ be a predicate. The set ${\cal A}_p = \{c_1,\ldots,c_n\}$
of symbolic conditions
over the integer arguments of $p$
is called {\em set of adornments for $p$} if $\bigvee_{i=1}^n c_i = \mbox{\sl true}$
and for all $i,j$ such that 
$1\leq i < j\leq n$, $c_i \wedge c_j = \mbox{\sl false}$.
\end{definition}

A set of adornments partitions the domain for (some of) the integer variables
of the predicate. Similarly to Example~\ref{example:osc},
in the examples to come, elements of a set of adornments  are denoted
{\mbox{\sf a}}, {\mbox{\sf b}}, {\mbox{\sf c}}, \ldots 

\begin{example}
\label{example:osc:cases}
Example~\ref{example:osc}, continued. The following
are examples of sets of adornments: 
\[
\begin{array}{ll}
\{{\mbox{\sf a}}, {\mbox{\sf b}}, {\mbox{\sf c}}\} &\mbox{\rm where}\;{\mbox{\sf a}}\;\mbox{\rm is}\;1 < p_1 < 1000\mbox{,}\;{\mbox{\sf b}}\;\mbox{\rm is}\; -1000 < p_1 < -1\\
& \mbox{\rm and}\;{\mbox{\sf c}}\;\mbox{\rm is}\; p_1 \leq -1000 \vee -1 \leq p_1 \leq 1 \vee p_1 \geq 1000\mbox{.}
\end{array}
\]
and
\[
\{{\mbox{\sf d}}, {\mbox{\sf e}}\}\;\mbox{\rm where}\;{\mbox{\sf d}}\;\mbox{\rm is}\;p_1 \leq 100\;\mbox{\rm and}\;{\mbox{\sf e}}\;\mbox{\rm is}\; p_1 > 100\mbox{.}
\]
$\hfill\Box$\end{example}

In the next section we are going to present a transformation, related to
the multiple specialisation technique. To define it formally we introduce 
the following definition:

\begin{definition}
\label{def:prefix}
Let $H\leftarrow B_1,\ldots,B_n$ be a rule. $B_1,\ldots,B_i$, 
is called {\em an integer prefix of the rule}, if for all $j$, 
$1\leq j\leq i\leq n$,
$B_j$ is an integer inequality and the only variables 
in its arguments are variables of $H$. $B_1,\ldots,B_i$
is called {\sl the maximal integer prefix} of the rule,  if it is an
integer prefix and 
$B_1,\ldots,B_i,B_{i+1}$ is not an integer prefix.
\end{definition}

Since an integer prefix constrains only variables appearing in the head
of a clause, there exists a symbolic condition over the
arguments of the predicate of the head, such that the 
integer prefix is its instance
with respect to the head. In general, this symbolic condition is 
not necessarily unique.
\begin{example}
Consider the following program: $p(X, Y, Y) \leftarrow Y > 5.$
The only integer prefix of this rule is $Y > 5$. There are two symbolic
conditions over the arguments of $p$, $p_2 > 5$ and $p_3 > 5$,
such that $Y > 5$ is their instance
with respect to $p(X, Y, Y)$. 
$\hfill\Box$\end{example}

In order to guarantee the uniqueness of such symbolic conditions
we require integer argument positions in the heads
of the rules to be occupied by distinct variables. 
For the sake of simplicity we assume {\em all} argument positions in the heads
of the rules to be occupied by distinct variables. 
Apt {\em et al.}~\cite{Apt:Marchiori:Palamidessi} call such
a rule {\em homogeneous}. Analogously, a logic program is called homogeneous
if all its clauses are homogeneous. 
Programs can be easily rewritten to a homogeneous form
(see~\cite{Apt:Marchiori:Palamidessi}). In the following we assume that
all programs are homogeneous.

\subsection{Program transformation}
\label{section:program:transformation}
The next question that should be answered is how the program
should be transformed given a set of adornments. After this transformation
$p^c(X_1,\ldots,X_n)$ will behave with respect to the transformed program
exactly as $p(X_1,\ldots,X_n)$ does, for all calls that satisfy the condition
$c$. Intuitively, we replace each call to the predicate $p$ 
in the original program by a number of possible calls in the transformed one.

Given a program $P$ and a set of possible adornments 
${\cal A} = \bigcup_{p\in P} {\cal A}_p$, 
the transformation is performed in a number of steps. Below we use 
Example~\ref{example:loop:to7} as a running example to illustrate the
different steps. Recall that it consists of only one clause
\[\begin{array}{l} 
 p(X) \leftarrow X < 7, X1\;\mbox{\sl is}\;X+1, p(X1)\mbox{.}
\end{array}\]
As set of adornments we use
\[
{\cal A}_p = \{{\mbox{\sf a}}, {\mbox{\sf b}}\},\;\mbox{\rm where}\;{\mbox{\sf a}}\;\mbox{\rm is}\;p_1 < 7\;\mbox{\rm and}\;{\mbox{\sf b}}\;\mbox{\rm is}\; p_1\geq 7\mbox{.}
\]

\begin{enumerate}
\item For each clause $r$ in $P$ and for each call $p(t_1,\ldots,t_n)$ to 
a recursive predicate $p$ occurring in $r$ add 
$\bigvee_{c\in {\cal A}_p} c(p(t_1,\ldots,t_n))$ before $p(t_1,\ldots,t_n)$. By 
Definition~\ref{def:set:of:adornments} the disjunction is {\sl true},
thus, the transformed program is equivalent to the original one.
In the example, the clause is transformed to
\[\begin{array}{l} 
 p(X) \leftarrow X < 7, X1\;\mbox{\sl is}\;X+1, (X1 < 7\;;\; X1\geq 7), p(X1)\mbox{.}
\end{array}\]
\item For each clause, such that the head of the clause,
say $p(t_1,\ldots,t_n)$, has a recursive predicate $p$, add
$\bigvee_{c\in {\cal A}_p} c(p(t_1,\ldots,t_n))$ as the first subgoal in
its body.  As for the previous step, the introduced call is equivalent
to {\sl true}, so that the transformation is obviously correct. In the example,
we obtain:
\[\begin{array}{l} 
 p(X) \leftarrow (X < 7\;;\; X\geq 7), X < 7, \\
 \hspace{1.5cm} X1\;\mbox{\sl is}\;X+1, (X1 < 7\;;\; X1\geq 7), p(X1)\mbox{.}
\end{array}\]
\item Next, moving to an alternative procedural interpretation of disjunction,
for each clause in which we introduced a disjunction in one of the previous
two steps, and for each such introduced disjunction
$\bigvee_{c\in {\cal A}_p} c(p(t_1,\ldots,t_n))$ we split these disjunctions,
introducing a separate clause for each disjunct. Thus, we apply the
transformation
\begin{eqnarray*} 
H &\leftarrow& B_1,\ldots,(A_1\; ;\;\ldots\; ;\;A_k), \ldots, B_n\mbox{.}
\end{eqnarray*}
to
\begin{eqnarray*} 
H &\leftarrow& B_1,\ldots,A_1, \ldots, B_n\mbox{.}\\
H &\leftarrow& B_1,\ldots,A_2, \ldots, B_n\mbox{.}\\
&& \vdots \\
H &\leftarrow& B_1,\ldots,A_k, \ldots, B_n\mbox{.}
\end{eqnarray*}
to each disjunction introduced in steps 1 and 2.

For our running example, we obtain four clauses:
\[\begin{array}{l} 
 p(X) \leftarrow X < 7, X < 7, X1\;\mbox{\sl is}\;X+1, X1 < 7, p(X1)\mbox{.} \\
 p(X) \leftarrow X < 7, X < 7, X1\;\mbox{\sl is}\;X+1, X1 \geq 7, p(X1)\mbox{.} \\
 p(X) \leftarrow X \geq 7, X < 7, X1\;\mbox{\sl is}\;X+1, X1 < 7, p(X1)\mbox{.} \\
 p(X) \leftarrow X \geq 7, X < 7, X1\;\mbox{\sl is}\;X+1, X1 \geq 7, p(X1)\mbox{.} 
\end{array}\]

Note  that, although this transformation is logically correct, it is
{\sl not} correct for Prolog programs with non-logical features. For 
instance, in the presence of ``cut'', it may produce a different 
computed answer set. Also, in the context of ``read'' or ``write'' calls,
the procedural behaviour may become very different. However, for purely 
logical programs with integer computations, both the declarative
semantics and the computed answer semantics are preserved. Likewise, the
termination properties are also preserved. Indeed, the transformation described can be seen 
as a repeated unfolding of $;$ using the following clauses:
\begin{eqnarray*}
&& ;(X,Y)\leftarrow X\mbox{.}\\
&& ;(X,Y)\leftarrow Y\mbox{.}
\end{eqnarray*}
It is well-known that unfolding cannot introduce infinite derivations~\cite{Bossi:Cocco}.
On the other hand, an infinite derivation of the original program can be easily mimicked
by the transformed program. 
\\$\;$\\
From here on we will restrict our attention to purely logical programs,
augmented with integer arithmetic. To prepare the next step in the 
transformation, note that, in the program resulting from step 3, for
each rule $r$ and for each recursive predicate $p$:
\begin{itemize}
\item if a call $p(t_1,\ldots,t_n)$ occurs in $r$, then it is immediately
preceded by some $c(p(t_1,$ $\ldots,t_n))$,
\item if an atom $p(t_1,\ldots,t_n)$ occurs as the head of $r$, then it is 
immediately followed by some $c(p(t_1,\ldots,t_n))$.
\end{itemize}
$\;$\\
Moreover, since the elements ${\cal A}_p$ partition the domain
(see Definition~\ref{def:set:of:adornments}),
conjuncts like $c_i(p(t_1,\ldots,t_n)), p(t_1,\ldots,t_n)$ and 
$c_j(p(t_1,\ldots,t_n)), p(t_1,\ldots,t_n)$ for $i\neq j$, are mutually
exclusive, as well as the initial parts of the rules, like 
\[p(t_1,\ldots,t_n)\leftarrow c_i(p(t_1,\ldots,t_n))\;\mbox{\rm and}\;
p(t_1,\ldots,t_n)\leftarrow c_j(p(t_1,\ldots,t_n)),\] $i\neq j$.
This means that we can now safely rename the different cases apart.
\item Replace each occurrence of $c(p(t_1,\ldots,t_n)), p(t_1,\ldots,t_n)$
in the body of the clause with  $c(p(t_1,\ldots,t_n)), p^c(t_1,\ldots,t_n)$
and each occurrence of a rule
\[p(t_1,\ldots,t_n)\leftarrow c(p(t_1,\ldots,t_n)), B_1,\ldots,B_n\]
with a corresponding rule
\[p^c(t_1,\ldots,t_n)\leftarrow c(p(t_1,\ldots,t_n)), B_1,\ldots,B_n\mbox{.}\]
In our example we get:
\[\begin{array}{l} 
 p^{\mbox{\sf a}}(X) \leftarrow X < 7, X < 7, X1\;\mbox{\sl is}\;X+1, X1 < 7, p^{\mbox{\sf a}}(X1)\mbox{.} \\
 p^{\mbox{\sf a}}(X) \leftarrow X < 7, X < 7, X1\;\mbox{\sl is}\;X+1, X1 \geq 7, p^{\mbox{\sf b}}(X1)\mbox{.} \\
 p^{\mbox{\sf b}}(X) \leftarrow X \geq 7, X < 7, X1\;\mbox{\sl is}\;X+1, X1 < 7, p^{\mbox{\sf a}}(X1)\mbox{.} \\
 p^{\mbox{\sf b}}(X) \leftarrow X \geq 7, X < 7, X1\;\mbox{\sl is}\;X+1, X1 \geq 7, p^{\mbox{\sf b}}(X1)\mbox{.} 
\end{array}\]
Because of the arguments presented above, the renaming is obviously correct,
in the sense that the LD-trees that exist for the given program and for the
renamed program are identical, except for the names of the predicates and 
for a number of failing 1-step derivations (due to entering clauses that fail 
in their guard in the given program). As a result, both the semantics (up to
renaming) and the termination behaviour of the program are preserved.
\item Remove all rules with a maximal integer 
prefix which is inconsistent, and remove
from the bodies of the remaining clauses all subgoals that are preceded by an 
inconsistent conjunction of inequalities.
In the example, both rules defining 
$p^{\mbox{\sf b}}$ are eliminated and we obtain:
\[\begin{array}{l} 
 p^{\mbox{\sf a}}(X) \leftarrow X < 7, X < 7, X1\;\mbox{\sl is}\;X+1, X1 < 7, p^{\mbox{\sf a}}(X1)\mbox{.} \\
 p^{\mbox{\sf a}}(X) \leftarrow X < 7, X < 7, X1\;\mbox{\sl is}\;X+1, X1 \geq 7, p^{\mbox{\sf b}}(X1)\mbox{.} 
\end{array}\]
Performing this step requires 
verifying the consistency of a set of constraints, a task that might be 
computationally expensive. Depending on the constraints the implementation 
of our technique 
is supposed to deal with, the programmer can either opt for more restricted
but potentially faster solvers, such as linear rational 
solver \cite{CLP:Manual}, or for more powerful but potentially slower ones,
such as mixed integer programminging solver \cite{ILOG}.
\item Replace each rule 
\[p^c(t_1,\ldots,t_n)\leftarrow c(p(t_1,\ldots,t_n)), B_1,\ldots,B_n\] 
by a rule \[p^c(t_1,\ldots,t_n)\leftarrow B_1,\ldots,B_n.\]
In the example we obtain:
\[
\begin{array}{l} 
p^{\mbox{\sf a}}(X) \leftarrow X < 7, X1\;\mbox{\sl is}\;X+1, X1 < 7, p^{\mbox{\sf a}}(X1).\\
p^{\mbox{\sf a}}(X) \leftarrow X < 7, X1\;\mbox{\sl is}\;X+1, X1 \geq 7, p^{\mbox{\sf b}}(X1). 
\end{array}\]
which is the {\sl adorned} program, $P^{\cal A}$
($P^{\{\mbox{\sf a}, \mbox{\sf b}\}}$ in our case). 
Note that this last step is only
correct if we also transform the set of original queries. Namely, given a 
single predicate set of original atomic queries $S$ for $P$ and
a set of adornments ${\cal A} = \bigcup_{p\in P} {\cal A}_p$,
the corresponding set of queries 
considered for $P^{\cal A}$ is 
$S^{\cal A} 
= \{c_1(Q)\wedge Q^{c_1},\ldots,c_n(Q)\wedge Q^{c_n}\mid Q\in S, 
\{c_1,\ldots,c_n\} = {\cal A}_{\mbox{\sl rel}(Q)}
\}$, where $Q^c$ denotes $p^c(t_1,\ldots,t_n)$ if $Q$ is $p(t_1,\ldots,t_n)$.
In our running example the set of queries is
$\{z < 7\wedge p^{\mbox{\sf a}}(z), z\geq 7\wedge p^{\mbox{\sf b}}(z)\mid z\;\mbox{\rm is an
integer}\}$.
\end{enumerate}

Before stating our results formally we illustrate the transformation
by a second example.
\begin{example}
\label{example:osc:adornments}
Example~\ref{example:osc}, continued. 
With the first set of adornments from Example~\ref{example:osc:cases}
we obtain $P^{\{\mbox{\sf a},\mbox{\sf b},\mbox{\sf c}\}}$:
\[\begin{array}{l}
 p^{\mbox{\sf a}}(X) \leftarrow X > 1, X < 1000, X1\;\mbox{\sl is}\;-X*X, \\
 \hspace{1.0cm} -1000 < X1, X1 < -1, p^{\mbox{\sf b}}(X1)\mbox{.} \\
 p^{\mbox{\sf a}}(X) \leftarrow X > 1, X < 1000, X1\;\mbox{\sl is}\;-X*X, \\
 \hspace{1.0cm} (X1 \leq -1000 ; (-1 \leq X1, X1 \leq 1) ; X1\geq 1000), p^{\mbox{\sf c}}(X1)\mbox{.} \\
 p^{\mbox{\sf b}}(X) \leftarrow X < -1, X > -1000,  X1\;\mbox{\sl is}\;X*X, \\
 \hspace{1.0cm} 1 < X1, X1 < 1000, p^{\mbox{\sf a}}(X1)\mbox{.} \\
 p^{\mbox{\sf b}}(X) \leftarrow X < -1, X > -1000,  X1\;\mbox{\sl is}\;X*X, \\
 \hspace{1.0cm} (X1 \leq -1000 ; (-1 \leq X1, X1 \leq 1) ; X1\geq 1000), p^{\mbox{\sf c}}(X1)\mbox{.}
\end{array}\]
The set of queries to be considered is
\begin{eqnarray*}
&\{&\\
&& z > 1\;\wedge\;z < 1000\;\wedge\;p^{\mbox{\sf a}}(z),\\
&& z > -1000\;\wedge\;z < -1\;\wedge\;p^{\mbox{\sf b}}(z),\\
&& (z \leq -1000\vee (z\geq -1 \wedge z\leq 1)\vee z\geq 1000)\wedge p^{\mbox{\sf c}}(z)\\
&\mid& z\;\mbox{\rm is an integer}\\
&\}&
\end{eqnarray*}
If the second set of adornments is used, the program $P^{\{\mbox{\sf d},\mbox{\sf e}\}}$ is obtained:
\eat{ 
$p^{\mbox{\sf d}}(X) \leftarrow X > 1, X < 1000, X1\;\mbox{\sl is}\;-X*X, p^{\mbox{\sf d}}(X1)\mbox{.} \\
p^{\mbox{\sf e}}(X) \leftarrow X > 1, X < 1000, X1\;\mbox{\sl is}\;-X*X, p^{\mbox{\sf d}}(X1)\mbox{.} \\
p^{\mbox{\sf d}}(X) \leftarrow X < -1, X > -1000, X1\;\mbox{\sl is}\;X*X, p^{\mbox{\sf d}}(X1)\mbox{.} \\ 
p^{\mbox{\sf d}}(X) \leftarrow X < -1, X > -1000, X1\;\mbox{\sl is}\;X*X, p^{\mbox{\sf e}}(X1)\mbox{.}\\$
} 
\[\begin{array}{l}
p^{\mbox{\sf d}}(X) \leftarrow X > 1, X < 1000, X1\;\mbox{\sl is}\;-X*X, X1 \leq 100, p^{\mbox{\sf d}}(X1)\mbox{.} \\
p^{\mbox{\sf e}}(X) \leftarrow X > 1, X < 1000, X1\;\mbox{\sl is}\;-X*X, X1 \leq 100, p^{\mbox{\sf d}}(X1)\mbox{.} \\
p^{\mbox{\sf d}}(X) \leftarrow  X < -1, X > -1000, X1\;\mbox{\sl is}\;X*X, X1 \leq 100, p^{\mbox{\sf d}}(X1)\mbox{.} \\ 
p^{\mbox{\sf d}}(X) \leftarrow X < -1, X > -1000, X1\;\mbox{\sl is}\;X*X, X1 > 100, p^{\mbox{\sf e}}(X1)\mbox{.} 
\end{array}\]

Analogously, the following is the set of the corresponding queries
\[\{z \leq 100\;\wedge\;p^{\mbox{\sf d}}(z),z > 100\;\wedge\;p^{\mbox{\sf e}}(z)\mid z\;\mbox{\rm is an integer}\}\]
$\hfill\Box$\end{example}
 
Formally, the following lemma holds:
\begin{lemma}
\label{lemma:pres}
Let $P$ be a definite pure logical program with integer computations, 
let $Q$ be an atomic query, 
let ${\cal A} = \cup {\cal A}_p$ be a 
set of adornments and let $c$ be an
adornment in ${\cal A}_{\mbox{\sl rel}(Q)}$. 
Let $P^{\cal A}$ be a program 
obtained as described above with respect to ${\cal A}$.
Then, $c$ is a termination condition for $Q$ 
with respect to $P$ if and only
if $P^{\cal A}$ LD-terminates with respect to 
$c(Q)\wedge Q^c$.
\end{lemma}
\begin{proof}
The construction of $P^{\cal A}$
implies that the LD-tree of $c(Q)\wedge Q^{c}$ with respect to $P^{\cal A}$
is isomorphic to the LD-tree of $c(Q)\wedge Q$ with respect to $P$, 
implying the theorem.
\end{proof}

Again, in practice we do not prove termination of a single query, but of
a single predicate set of queries. Furthermore, recalling that a disjunction
of termination conditions is a termination condition itself we can generalise
our lemma to disjunctions of adornments. Taking these two considerations into
account, the following theorem holds.
\begin{theorem}
\label{theorem:pres}
Let $P$ be a definite pure logical program with integer computations, 
let $S$ be a single predicate set of atomic queries, 
let ${\cal A} = \cup {\cal A}_p$ be a 
set of adornments and let $c_1,\ldots,c_n$ be 
adornments in ${\cal A}_{\mbox{\sl rel}(S)}$. 
Let $P^{\cal A}$ be a program 
obtained as described above with respect to ${\cal A}$.
Then, $c_1\vee \ldots\vee c_n$ is a termination condition for $S$ 
with respect to $P$ if and only
if $P^{\cal A}$ LD-terminates with respect to 
$\{c_1(Q)\wedge Q^{c_1},\ldots,c_n(Q)\wedge Q^{c_n}\mid Q\in S\}$.
\end{theorem}
\begin{proof}
Immediately from Lemma~\ref{lemma:pres} and the preceding observations.
\end{proof}

The goal of the transformation presented is, given a program
and a partition of the domain, to generate a program having
separate clauses for each one of the cases. Clearly, this may
(and usually will) increase the number of clauses. Each clause can
be replaced by maximum $c^{n+1}$ new clauses, were $c$ is a number
of adornments and $n$ is a number of recursive body subgoals.
Thus, the size of the transformed program doesn't exceed 
\begin{eqnarray}
&& r\times c^{n+1}, \label{size:1}
\end{eqnarray} 
where $r$
is the size of the original one. This may seem a problematically
large increase, however, the number of recursive body atoms (depending on
numerical arguments) in numerical programs is usually small.

Since the transformation preserves
termination, acceptability of the transformed program implies
termination of the original program. In the next section we
will see that having separate clauses for different cases
allows us to define less sophisticated level-mappings for 
proving termination. Such level-mappings can be
constructed automatically, and thus, play a key role in automation of
the approach.

\section{Generating adornments, level mappings and termination constraints}
\label{section:generating}
In the previous section we have shown the transformation that allows
reasoning on termination of the numerical computations in the framework
of acceptability with respect to a set of queries. In this section we discuss
how adornments can be generated, how level mappings can be proposed and 
which termination conditions finally turn up.

\subsection{Guard-tuned sets of adornments}
\label{subsection:inferring}
In Example~\ref{example:osc:cases} we have seen two different sets of 
adornments. Both of them are valid according to Definition~\ref{def:set:of:adornments}. 
However, recalling $P^{\{\mbox{\sf a}, 
\mbox{\sf b},\mbox{\sf c}\}}$ and $P^{\{\mbox{\sf d},\mbox{\sf e}\}}$ 
as shown in 
Example~\ref{example:osc:adornments}, we conclude that $\{\mbox{\sf a}, 
\mbox{\sf b},\mbox{\sf c}\}$ is 
in some sense preferable to $\{\mbox{\sf d},\mbox{\sf e}\}$. 
Observe that $P^{\{\mbox{\sf d},\mbox{\sf e}\}}$ does not only have two 
mutually
recursive predicates, as $P^{\{\mbox{\sf a}, \mbox{\sf b},\mbox{\sf c}\}}$ 
does, but also self-loops on one of the predicates. 
To distinguish between ``better'' and ``worse'' sets of adornments
we define {\em guard-tuned} sets of adornments.

Intuitively, a set of adornments of a predicate $p$ is 
{\em guard-tuned} if it is based on ``subcases'' of maximal 
integer prefixes.

\begin{definition}
Let $P$ be a homogeneous program, 
let $p$ be a predicate in $P$. A set of adornments ${\cal A}_p$
is called {\em guard-tuned\/}
if for every $A\in {\cal A}_p$ and
for every rule $r\in P$, defining $p$, with the symbolic condition $c$
corresponding to its maximal integer 
prefix,  either $c \wedge A = \mbox{\sl false}$
or $c \wedge A = A$ holds.
\end{definition}

\begin{example}
\label{example:osc:cases2}
The first set of adornments, presented in 
Example~\ref{example:osc:cases}, is guard-tuned while
the second one is not guard-tuned.
$\hfill\Box$\end{example}

Examples~\ref{example:osc:cases} and \ref{example:osc:cases2}
suggest the following way of constructing  a guard-tuned set of adornments.
\eat{
Given a program $P$ one might collect the symbolic 
conditions, corresponding to the maximal 
integer prefixes of the rules defining
a predicate $p$ 
(we denote this set ${\cal C}_p$) and
add the ``completion'' of the constructed disjunction. 
Here, with ``completion'' we mean adding cases in order to cover the
full domain of integers.
Unfortunately, this set 
is not necessarily a set of adornments and if so, it
is not necessary guard-tuned.

\begin{example}
\label{example:intersections:needed}
Consider the following program.
\[\begin{array}{l}
 r(X) \leftarrow X > 5\mbox{.}\;\;r(X) \leftarrow X > 10, r(X)\mbox{.}
\end{array}\]
Two sets of symbolic conditions can be constructed:
$\{r^{\$1\leq 5}, r^{\$1 > 5}, r^{\$1 > 10}\}$ which is not a set of adornments
and $\{r^{\$1\leq 5}, r^{\$1 > 5}\}$ which is not guard-tuned.
$\hfill\Box$\end{example}
We use a different approach.
}
First, we collect the symbolic 
conditions, corresponding to the maximal integer
prefixes of the rules defining
a predicate $p$ (we denote this set ${\cal C}_p$). Let ${\cal C}_p$ be
$\{c_1, \ldots, c_n\}$. Then we define ${\cal A}_p$ to be the set of all
conjunctions $\wedge^n_{i=1} d_i$, where $d_i$ is either $c_i$ or $\neg c_i$. 
Computing ${\cal A}_p$ might be exponential in the number of elements
of ${\cal C}_p$, i.e., in the number of maximal 
integer prefixes. The number of integer prefixes is 
bounded by the number of clauses. Thus, recalling (\ref{size:1}), the upper
bound on the size of the transformed program is $r\times 2^{r(n+1)}$,
i.e., it is exponential in the number of clauses $r$ and in a number of
recursive subgoals $n$. However, again our experience suggests that numerical 
parts of real-world programs are usually relatively small and depend on
one or two different integer prefixes. Analogously, clauses having more
than two recursive body subgoals are highly exceptional. Therefore, we
conclude that in practice the size of the transformed program is not 
problematic.

We claim that the constructed set ${\cal A}_p$ is
always a guard-tuned set of adornments. Before stating this formally, consider
the following example.
\begin{example}
Consider the following program.
\begin{eqnarray*}
&& r(X) \leftarrow X > 5\mbox{.}\\
&& r(X) \leftarrow X > 10, r(X)\mbox{.}
\end{eqnarray*}

Then, ${\cal C}_r = \{r_1 > 5, r_1 > 10\}$. The following
conjunctions can be constructed from the elements of ${\cal C}_p$ and their
negations:
$\{r_1 > 5\wedge r_1 > 10,\;r_1 > 5 \wedge \neg(r_1 > 10),\;\neg(r_1 > 5) \wedge r_1 > 10,\;\neg(r_1 > 5)\wedge \neg(r_1 > 10)\}$.
After simplifying and removing inconsistencies
${\cal A}_r = \{r_1 > 10, r_1 > 5 \wedge r_1 \leq 10, r_1 \leq 5\}$.
$\hfill\Box$\end{example}
\begin{lemma}
\label{lemma}
Let $P$ be a program, $p$ be a predicate in $P$ and ${\cal A}_p$ be 
constructed as described. Then ${\cal A}_p$ is a guard-tuned set of adornments.
\end{lemma}
\begin{proof*}
The proof is done by checking the definitions.
\begin{enumerate}
\item Let $a_1, a_2\in{\cal A}_p$ and $a_1\neq a_2$. Then, there exists $c_i\in {\cal C}_p$,
such that $a_1 = d_1\wedge\ldots\wedge c_i \wedge\ldots\wedge d_n$ and $a_2 = d_1\wedge\ldots\wedge \neg c_i \wedge\ldots\wedge d_n$. Thus,
$a_1\wedge a_2 = \mbox{\sl false}$. 
\item By definition of ${\cal A}_p$, $\vee_{a_i\in {\cal A}_p} a_i = \mbox{\sl true}$. Thus,
${\cal A}_p$ is a set of adornments.
\item Let $a\in {\cal A}_p$ be an adornment and let $c$ be a symbolic condition 
corresponding to the maximal integer 
prefix of a rule. By definition of ${\cal C}_p$,
$c\in {\cal C}_p$. Thus, either $c$ is one of the conjuncts of $a$ or $\neg c$ 
is one of the conjuncts of $a$. In the first case, $c\wedge a = a$. In the second
case $c\wedge a = c\wedge (\neg c) = \mbox{\sl false}$. Therefore, ${\cal A}_p$
is a guard-tuned set of adornments.
\end{enumerate}
$\hfill\blacksquare$\end{proof*}

From here on we assume that all sets of adornments are guard-tuned.

\subsection{How to define a level mapping.}
\label{subsection:lm}
One of the questions that should be answered is how the level mappings
should be generated automatically. Clearly, one cannot expect automatically
defined level mappings to be powerful enough to prove termination of all terminating
examples. In general we cannot hope but for a good guess.

The problem with level mappings is that they should reflect changes
on possibly negative arguments and remain
non-negative at the same time. We also like to remain in the
framework of level mappings on atoms defined
as linear combinations of sizes of their arguments~\cite{Bossi:Cocco:Fabris}.
We solve this problem by defining different level mappings
for different adorned versions of the predicate.
The major observation underlying
the technique presented in this subsection is that if $p_1 > p_2$
appears in the adornment of a recursive clause, 
then for each call to this adorned predicate
$p_1 - p_2$ will be positive, and thus, can be used for defining a 
level mapping. On the other hand, $p_1 < p_2$ can always be interpreted
as $p_2 > p_1$. These observations form a basis for definition of   
a {\em primitive level mapping}.
\begin{definition}
\label{def:primitive:lm}
Let $p^c$ be an adorned predicate. 
The {\em primitive level mapping}, $\mid \cdot \mid^{\mbox{\sl pr}}$,
is defined as 
\begin{itemize}
\item if $c$ is $E_1\;\rho\;E_2$, where
$E_1$ and $E_2$ are expressions and $\rho$ is either $>$ or $\geq$ then
\[\!\!\!\!\!\!\!\!\mid p^{c}(t_1,\ldots,t_n)\mid^{\mbox{\sl pr}} = 
\left\{
\begin{array}{ll}
(E_1 - E_2)(t_1,\ldots,t_n) & \mbox{\rm if}\;\;E_1(t_1,\ldots,t_n)\;\rho\;E_2(t_1,\ldots,t_n) \\
0         & \mbox{\rm otherwise}
\end{array}
\right.\] 
\item if $c$ is $E_1\;\rho\;E_2$, where
$E_1$ and $E_2$ are expressions and $\rho$ is either $<$ or $\leq$ then
\[\!\!\!\!\!\!\!\!\mid p^{c}(t_1,\ldots,t_n)\mid^{\mbox{\sl pr}} = 
\left\{
\begin{array}{ll}
(E_2 - E_1)(t_1,\ldots,t_n) & \mbox{\rm if}\;\;E_1(t_1,\ldots,t_n)\;\rho\;E_2(t_1,\ldots,t_n) \\
0         & \mbox{\rm otherwise}
\end{array}
\right.\] 
\item otherwise,  
\[\!\!\!\!\!\!\!\!\mid p^{c}(t_1,\ldots,t_n)\mid^{\mbox{\sl pr}} = 0.\]
\end{itemize}
\eat{
\begin{itemize}
\item if $c = E_1\;\rho\;E_2$, where
$E_1$ and $E_2$ are expressions and $\rho\in\{>,\geq\}$, then
\[\left\{
\begin{array}{ll}
(E_1 - E_2)(t_1,\ldots,t_n) & \mbox{\rm if}\;\;E_1(t_1,\ldots,t_n)\;\rho\;E_2(t_1,\ldots,t_n) \\
0         & \mbox{\rm otherwise}
\end{array}
\right. \]
\item $0$, otherwise.
\end{itemize}
} 
\end{definition}

If more than one conjunct appears in the 
adornment, the level mapping is defined as a linear 
combination of primitive level mappings corresponding to the conjuncts.

\begin{definition}
Let $p^{c_1\wedge\ldots\wedge c_n}$ be an adorned predicate
such that each $c_i$ is $E^i_1\rho^i E^i_2$
for some expressions $E^i_1$ and $E^i_2$ and $\rho^i$ is either
$>$ or $\geq$. Let $w_{c_1},\ldots,w_{c_n}$ be natural numbers. Then, a 
level mapping $\mid \cdot\mid$ satisfying
\[\mid p^{c_1\wedge\ldots\wedge c_n}(t_1,\ldots,t_n) \mid\;\; =
\sum_{i} w_{c_i} \mid p^{c_i}(t_1,\ldots,t_n) \mid^{\mbox{\sl pr}},\] 
is called a {\em natural level mapping}.
\end{definition}
\begin{example}
\label{example:osc:nat:lm}
The level mappings used in Example~\ref{example:osc} are natural
level mappings such that
$w_{p_1 > 1} = w_{p_1 < -1} = 0$, $w_{p_1 < 1000} = w_{p_1 > -1000} = 1$.
We have seen that these level mappings are powerful enough to prove 
termination.
$\hfill\Box$\end{example} 

The definition of natural level mapping implies that if $c$ is a disjunction, it is ignored. The reason for doing so is that disjunctions 
are introduced only as negations of symbolic constraints corresponding to maximal integer prefixes of the rules. Thus, they signify that some rule {\em cannot} be applied, and can be ignored.

\begin{example}
Example~\ref{example:osc}, continued. Recalling that 
${\mbox{\sf c}}$ denotes $p_1 \leq -1000 \vee -1 \leq p_1 \leq 1 \vee p_1 \geq 1000$
the following holds for any integer $n$,
$\mid p^{\mbox{\sf c}}(n)\mid^{\mbox{\sl pr}}\;\; = 0$.
$\hfill\Box$\end{example}

Of course, if the original program already contains disjunctions of numerical 
constraints, then we transform it in a preprocessing to eliminate the 
disjunctions.

As the following example illustrates, natural level mappings gain their power 
from the fact that the set of adornments used is guard-tuned. 

\begin{example}
In Example~\ref{example:osc:cases} we have seen two different sets of 
adornments. We have seen in Example~\ref{example:osc} 
that if a guard-tuned set of adornments is chosen 
the natural level mapping is powerful enough to prove acceptability and, thus,
termination. If a non guard-tuned set of adornments is chosen, the second 
program of Example~\ref{example:osc:adornments} is obtained. Then, a following 
natural level mapping is defined (for some natural numbers 
$w_{p_1 \leq 100}$ and $w_{p_1 > 100}$):
\[\begin{array}{lcl}
\mid p^{\mbox{\sf d}}(X)\mid &=& w_{p_1 \leq 100} * \left\{
  \begin{array}{ll}
  100 - X & \mbox{\rm if $X \leq 100$}\\
  0       & \mbox{\rm otherwise}
  \end{array}
 \right.\\
\mid p^{\mbox{\sf e}}(X)\mid &=& w_{p_1 > 100} * \left\{
  \begin{array}{ll}
  X - 100 & \mbox{\rm if $X > 100$}\\
  0       & \mbox{\rm otherwise}
  \end{array}
 \right. 
\end{array}
\]
Consider the following clause.
\begin{eqnarray*} 
&& p^{\mbox{\sf d}}(X) \leftarrow X > 1, X < 1000, X1\;\mbox{\sl is}\;-X*X, X1 \leq 100, p^{\mbox{\sf d}}(X1)\mbox{.} 
\end{eqnarray*}
In order to prove acceptability we have to show that the size of the call
to $p^{\mbox{\sf d}}(X)$ is greater than the size of the corresponding call 
to $p^{\mbox{\sf d}}(X1)$. If the first argument $x$ at the call to 
$p^{\mbox{\sf d}}(X)$ is greater than 1 and less than 10, the acceptability
decrease requires $w_{p_1 \leq 100}(100 - x) > w_{p_1 \leq 100}(100 + x^2)$, contradicting $x > 1$ and $w_{p_1 \leq 100}$ being a natural number. Thus,
acceptability cannot be proved with natural level mappings.
$\hfill\Box$\end{example}

The approach of~\cite{Decorte:DeSchreye:Vandecasteele} 
defines symbolic counterparts of the level
mappings and infers the values of the coefficients by solving
a system of constraints. Intuitively, instead of considering $w_{c_i}$'s as
given coefficients, they are regarded as variables. More formally,
similarly to~\cite{Decorte:DeSchreye:Vandecasteele}, we introduce the following notion.

\begin{definition} 
\label{def:s:c}
Let $p^{c_1\wedge\ldots\wedge c_n}$ be an adorned predicate. 
A {\em symbolic counterpart of a natural level mapping}
is an expression: \[\mid p^{c_1\wedge\ldots\wedge c_n}(t_1,\ldots,t_n) \mid^s\;\; =
\sum_{i} W_{c_i} \mid p^{c_i}(t_1,\ldots,t_n) \mid^{\mbox{\sl pr}},\] 
where the $W_{c_i}$'s are symbols, associated to a predicate $p^{c_1\wedge\ldots\wedge c_n}$.
\end{definition}

The intuition behind the symbolic counterpart of a natural level mapping
is that natural level mappings are instances of it. Therefore, we also
require $W_c\geq 0$ to hold for any constraint $c$.

\begin{example}
\label{example:sym}
Example~\ref{example:osc}, continued.
Recalling that ${\mbox{\sf a}}$ stands for $1 < p_1 < 1000$,
a symbolic counterpart of a natural level mapping for $p^{\mbox{\sf a}}(n)$ 
is $W_{p_1 > 1}(n-1) + W_{p_1 < 1000}(1000-n)$.
$\hfill\Box$\end{example} 

In order to verify the rigid acceptability condition (Theorem~\ref{rigid:acc:term}) interargument relations may be required as well. Interargument relations are usually represented as saying that a weighted sum of sizes of some arguments (with respect to a given norm) is greater or equal to a weighted sum of sizes of other arguments (see e.g.~\cite{Plumer:ICLP}). In the numerical case these sizes should be replaced with expressions as used in Definition~\ref{def:primitive:lm}. Observe that for simpler examples no interargument relations are needed. Symbolic counterparts of norms and interargument relations can be defined analogously to Definition~\ref{def:s:c}. In the next subsection we discuss how the symbolic counterparts are used to infer termination conditions.

\subsection{Inferring termination constraints}
In this section, we combine the steps studied so far into 
an algorithm that infers termination conditions. The program transformation,
described in Section~\ref{section:program:transformation}, implies
that a trivial termination condition can be computed as a disjunction
of the adornments corresponding to the predicates that can be completely
unfolded, i.e., to the predicates that do not depend directly or indirectly on
recursive predicates. More formally we can draw the following corollary from 
Theorem~\ref{theorem:pres}: 
\begin{corollary}
\label{cor:nonrec}
Let $P$ be a program, let $S$ be a single predicate set of atomic queries
and let ${\cal A}$ be a set of adornments for $\mbox{\sl rel}(S)$. Let 
$A = \{c\mid c\in {\cal A}, \mbox{\rm for all}\;q\;\mbox{\rm such that}\;\mbox{\sl rel}(S)^c \sqsupseteq q:\;q\;\mbox{\rm is not recursive}$ $\mbox{\rm in}
 P^{\cal A}\}$. Then $\bigvee_{c\in A} c$ is a termination condition for $S$
with respect to $P$.
\end{corollary}
\begin{proof*}
By definition of $A$, for all $Q\in S$, 
$P^{\cal A}$ 
LD-terminates with respect to $\{c(Q)\wedge Q^c \mid c\in A\}$. Thus,
by Theorem~\ref{theorem:pres}, $\bigvee_{c\in A} c$ is a termination condition
for $S$ with respect to $P$.
$\hfill\blacksquare$\end{proof*}

\begin{example}
The termination condition constructed according to  Corollary~\ref{cor:nonrec}
for Example~\ref{example:osc} is ${\mbox{\sf c}}$, i.e.,
$(p_1\leq -1000) \vee (-1 \leq p_1 \leq 1) \vee (p_1\geq 1000)$.
$\hfill\Box$\end{example}

In general, the termination condition is constructed as a 
disjunction of two conditions: $\mbox{\sl cond}_1$ for non-recursive cases, according to 
Corollary~\ref{cor:nonrec}, and $\mbox{\sl cond}_2$, for recursive cases.
The later condition is initialised to be $\neg \mbox{\sl cond}_1$ and further
refined by adding constraints obtained from the rigid acceptability condition, 
as in~\cite{Decorte:DeSchreye:Vandecasteele}\footnote{
Any other technique proving termination and
able to 
provide 
some constraint that, if satisfied, implies termination can be 
used instead of~\cite{Decorte:DeSchreye:Vandecasteele}.}.
The algorithm is sketched in Figure~\ref{algo}. 

Termination inference is inspired by the
constraints-based approach of Decorte {\em et al.}~\cite{Decorte:DeSchreye:Vandecasteele}. Similarly to their work we start by constructing symbolic counterparts of the level mappings (Definition~\ref{def:s:c}) and interargument 
relations, and construct conditions following from rigid 
acceptability (Theorem~\ref{rigid:acc:term}) and validity of 
interargument relations. Unlike their work in our case no rigidity
constraints are needed (since integer arguments are ground and obviously 
rigid) and norms are fixed. Thus, the constraints system 
turns out to be simpler and better suited for automation. Finally, 
the conditions constructed are solved with respect to the symbolic 
variables ($W_{c_i}$'s). In Example~\ref{example:gcd:simplified:2} below we
are going to see that rigid acceptability will be implied by the following
constraint ($\Sigma$):
\begin{eqnarray*}
&& W_{q_1 > q_2}(X-Y) > W_{q_1 > q_2}((X-Y)-Y), 
\end{eqnarray*}
that is $W_{q_1 > q_2} Y > 0$ should hold. In both approaches
$W_{q_1 > q_2}\geq 0$ is required to hold. At this point 
the approach of~\cite{Decorte:DeSchreye:Vandecasteele}, interpreting 
$Y$ as a norm of arguments (i.e., $Y$ is a natural number), 
will conclude $W_{q_1 > q_2} > 0$. In our case, we do not know {\em a priori} 
that $Y$ is a natural number. Therefore, we would infer from
$\Sigma$ that $Y > 0$ and 
$W_{q_1 > q_2} > 0$.

In general, given the system of constraints
inferred by~\cite{Decorte:DeSchreye:Vandecasteele},
we distinguish between the following cases:
\begin{itemize}
\item There is no solution. We report $\mbox{\sl cond}_1$ as a termination condition (Corollary~\ref{cor:nonrec}). Observe that when the algorithm reports the termination condition to be {\sl false},
it suspects the possibility of non-termination.
\item There is a solution for any values of integer variables. Namely,
there are natural level mappings and interargument relations that prove 
termination of the program for {\em any} values of integer variables.
Termination condition in this case is, thus, {\em true}.
\item There is a solution for some values of integer variables. In other
words, the solution constrains integer variables appearing in the clauses.
Two cases can be distinguished:
\begin{itemize}
\item Integer variables constrained appear in the heads of the clauses.
Then, constraints on these variables can be regarded as constraints on
the arguments of the queries posed. In this case termination can be shown
if these constraints are fulfilled.
\item Integer variables constrained do not appear in the heads of the clauses.
In this case our methodology is too weak to obtain some information implying
termination of the queries. The best we can do is to report termination
for $\mbox{\sl cond}_1$.
\end{itemize}
\end{itemize}

\begin{figure*}[!ht]
\begin{center}
\fbox{
\parbox{4.6in}{
Let $P$ be a homogeneous
program, let $S$ be a single predicate set of atomic queries 
and let $q$ be $\mbox{\sl rel}(S)$.
\begin{enumerate}
\item For each $p\simeq q$ construct a guard-tuned set ${\cal A}_p$. (Section~\ref{subsection:inferring})
\item Adorn $P$ with respect to $q$ and $\bigcup_{p\simeq q}{\cal A}_p$. (Section~\ref{section:program:transformation})
\item Let $A = \{c\mid c\in {\cal A}_q, \mbox{\sl for all}\;p\;\mbox{\sl such that}\;q^c \sqsupseteq p\;:\;p\;\mbox{\sl is not recursive in $P^{\cal A}$}\}$. \\ Let  $\mbox{\sl cond}_1 = \bigvee_{c\in A} c$. Let  $\mbox{\sl cond}_2 = \bigvee_{c\in {\cal A}_q, c\not\in A} c$.
\item Let $S'$ be $\{c(Q)\wedge Q^c\mid c\in {\cal A}_q, c\not\in A, c(Q)\wedge Q^c\in S^{\cal A}\}$.
\item Define the symbolic counterparts of level mappings and interargument relations. (Section~\ref{subsection:lm})
\item Let $\Sigma$ be a set of constraints on the symbolic variables, following from rigid acceptability of $S'$ with respect to $P^{\cal A}$ 
and validity of interargument relations.
\item Solve $\Sigma$ with respect to the symbolic variables.
\begin{enumerate}
\item Solution of $\Sigma$ doesn't produce extra constraints on variables.
\begin{itemize}
\item[] Report termination for $\mbox{\sl true}$.
\end{itemize}
\item Solution of $\Sigma$ produces extra constraints on
integer variables, appearing in the heads of the clauses.
\begin{itemize}
\item[] Conjunct these constraints to termination condition $\mbox{\sl cond}_2$.
\item[] Report termination for $\mbox{\sl cond}_1 \vee \mbox{\sl cond}_2$.
\end{itemize}
\item There is no solution or integer variables, constrained by the 
solution of $\Sigma$, do not appear in the heads of the clauses
\begin{itemize}
\item[] Report termination for $\mbox{\sl cond}_1$.
\end{itemize}
\end{enumerate}
\end{enumerate}
}}
\caption{Termination Inference Algorithm}
\label{algo}
\end{center}
\end{figure*}
\begin{example}
\label{example:gcd:simplified:2}
Consider the following program.
\[\begin{array}{l}
 q(X,Y) \leftarrow X>Y, Z\;\mbox{\sl is}\;X-Y, q(Z,Y).
\end{array}\]
We look for integer values of $X$ and $Y$ such that $q(X,Y)$
terminates. First, the algorithm infers adornments. In our case
$\{{\mbox{\sf a}}, {\mbox{\sf b}}\}$ are inferred, such that ${\mbox{\sf a}}$ denotes
$q_1>q_2$ and ${\mbox{\sf b}}$ denotes $q_1\leq q_2$.

The adorned version of this program is 
\[\begin{array}{l}
 q^{\mbox{\sf a}}(X,Y) \leftarrow X>Y, Z\;\mbox{\sl is}\;X-Y, Z > Y, q^{\mbox{\sf a}}(Z,Y).\\
 q^{\mbox{\sf a}}(X,Y) \leftarrow X>Y, Z\;\mbox{\sl is}\;X-Y, Z \leq Y, q^{\mbox{\sf b}}(Z,Y).
\end{array}\]
The corresponding set of queries is 
\[\{x > y\wedge q^{\mbox{\sf a}}(x,y),
x\leq y\wedge q^{\mbox{\sf b}}(x,y)\mid x,y\;\mbox{\rm are integers}\}.\]

There is no clause defining $q^{\mbox{\sf b}}$. By Corollary~\ref{cor:nonrec}, ${\mbox{\sf b}}$, 
i.e., $q_1\leq q_2$ is a termination condition. This is the one 
we denoted $\mbox{\sl cond}_1$. The termination condition for  $q^{\mbox{\sf a}}$, denoted
$\mbox{\sl cond}_2$, is initialised to be $q_1 > q_2$.
The symbolic counterpart of a natural level mapping is
\[\mid q^{\mbox{\sf a}}(X,Y) \mid\;= W_{q_1 > q_2} * \left\{ \begin{array}{ll}
X-Y & \mbox{\rm if}\;\;X>Y \\
0 & \mbox{\rm otherwise}\end{array} \right.\]

The set of constraints $\Sigma$ implied by rigid acceptability is:
\begin{eqnarray}
&& W_{q_1 > q_2}(X-Y) > W_{q_1 > q_2}((X-Y)-Y), \label{example:constr}
\end{eqnarray}
that is $W_{q_1 > q_2} Y > 0$ should hold.
Since $W_{q_1>q_2} \geq 0$, $Y>0$ and
$W_{q_1 > q_2} > 0$ should hold. Variable $Y$ appears in the head of the
clause, i.e., $Y > 0$ can be viewed as a constraint on the query.
We update $\mbox{\sl cond}_2$ to be $(q_1 > q_2) \wedge (q_2 > 0)$
and report termination for $q_1\leq q_2 \vee (q_1 > q_2 \wedge q_2 > 0)$.
\eat{
Now we restart the entire process with respect to $Y>0$. 
Adornments inferred are $q_1 > q_2 \wedge q_2 > 0$ (denoted ${\mbox{\sf c}}$),
$q_1 > q_2 \wedge q_2 \leq 0$ (denoted ${\mbox{\sf d}}$),
$q_1 \leq q_2 \wedge q_2 > 0$ (denoted ${\mbox{\sf e}}$), and
$q_1 \leq q_2 \wedge q_2 \leq 0$ (denoted ${\mbox{\sf f}}$).

The following adorned program is obtained:
\[\begin{array}{l}
 q^{\mbox{\sf c}}(X,Y) \leftarrow X>Y, Y>0, Z\;\mbox{\sl is}\;X-Y,\\
 \hspace{2.0cm}  Z>Y, Y>0, q^{\mbox{\sf c}}(Z,Y).\\
 q^{\mbox{\sf d}}(X,Y) \leftarrow X>Y, Y\leq 0, Z\;\mbox{\sl is}\;X-Y,\\
 \hspace{2.0cm}   Z>Y, Y\leq 0, q^{\mbox{\sf d}}(Z,Y).\\
 q^{\mbox{\sf d}}(X,Y) \leftarrow X>Y, Y\leq 0, Z\;\mbox{\sl is}\;X-Y, \\
 \hspace{2.0cm} Z\leq Y, Y\leq 0, q^{\mbox{\sf e}}(Z,Y).
\end{array}\]
The corresponding set of queries is
\begin{eqnarray*}
&&\{x > y\;\wedge\;y > 0\;\wedge\;p^{\mbox{\sf c}}(x,y), x > y\;\wedge\;y \leq 0\;\wedge\;p^{\mbox{\sf d}}(x,y),\\
&& x \leq y\;\wedge\;y > 0\;\wedge\;p^{\mbox{\sf e}}(x,y), x \leq y\;\wedge\;y \leq 0\;\wedge\;p^{\mbox{\sf f}}(x,y)  \mid x,y\;\mbox{\rm are integers}\}
\end{eqnarray*}

\eat{ 
$q^{\$1>\$2,\$2>0}(X,Y) \leftarrow X>Y, Z\;\mbox{\sl is}\;X-Y, q^{\$1>\$2,\$2>0}(Z,Y).\\
q^{\$1>\$2,\$2\leq 0}(X,Y) \leftarrow X>Y, Z\;\mbox{\sl is}\;X-Y, q^{\$1>\$2,\$2 \leq 0}(Z,Y).\\
q^{\$1>\$2,\$2\leq 0}(X,Y) \leftarrow X>Y, Z\;\mbox{\sl is}\;X-Y, q^{\$1\leq \$2,\$2 \leq 0}(Z,Y).
$ } 

The second and the third clauses are 
removed, since they are ``irrelevant'' with respect to $q_2 > 0$.  
Thus, the only clause that should be analysed is the
first one. The level mapping is defined as
\[\mid q^{\mbox{\sf c}}(X,Y) \mid\;= W_{q_1 > q_2} * \left\{ \begin{array}{ll}
X-Y & \mbox{\rm if}\;\;X>Y \\
0 & \mbox{\rm otherwise}\end{array} \right. +
W_{q_2 > 0} * \left\{ \begin{array}{ll}
Y & \mbox{\rm if}\;\;Y>0 \\
0 & \mbox{\rm otherwise}\end{array} \right.\]

Rigid acceptability decreases imply 
\[W_{q_1 > q_2} (X-Y) + W_{q_2 > 0} Y > W_{q_1 > q_2} ((X-Y) - Y) + W_{q_2 > 0} Y,\] 
i.e., $0 > - W_{q_1 > q_2} Y$. This inequality holds, since
$Y > 0$ and $W_{q_1 > q_2} > 0$ are assumed to hold.
This solution does not impose
additional constraints on the integer variables. Thus, the analysis terminates
reporting $q_1\leq q_2 \vee (q_1 > q_2 \wedge q_2 > 0)$ 
as a termination condition.
}
$\hfill\Box$\end{example}

Formally the following theorem holds.
\begin{theorem}
Let $P$ be a homogeneous pure logical program with integer computation,
let $S$ be a single predicate set of atomic queries and let {\tt Algo} be the algorithm presented in Figure~\ref{algo}.
Then the following holds:
\begin{itemize}
\item $\mbox{\tt Algo}(P,S)$ terminates;
\item Let $c$ be a symbolic condition returned by $\mbox{\tt Algo}(P,S)$. Then 
$c$ is a termination condition for $S$.
\end{itemize}
\end{theorem}
\begin{proof*}
\begin{itemize}
\item Termination of $\mbox{\tt Algo}(P,S)$ follows from termination of its steps. Termination of steps 1 and 2 follows from the presentation of these transformations in Sections~\ref{subsection:inferring} and~\ref{section:program:transformation}, respectively. Termination of steps 3--7 is obvious.
\item Partial correctness follows from the correctness of transformations
and the corresponding result of~\cite{Decorte:DeSchreye:Vandecasteele}.
Correctness of step 1 is established by Lemma~\ref{lemma}, of step 2 by 
Theorem~\ref{theorem:pres}. For step 4 observe that termination for
queries in $S^{\cal A}\backslash S'$ is obvious by choice of $A$.
Correctness of steps 6 and 7 follows from 
the corresponding result of~\cite{Decorte:DeSchreye:Vandecasteele}.
\end{itemize}
$\hfill\blacksquare$\end{proof*}

In Example~\ref{example:gcd:simplified:2} the termination condition inferred by
our algorithm was optimal, i.e., any other termination condition implies it.
However, undecidability of the termination problem implies that no automatic 
tool can always guarantee optimality.
\begin{example}
\label{example:gcd:simplified:2a}
Consider the following program.
\begin{eqnarray*}
&& q(X,Y) \leftarrow X > Y, Z\;\mbox{\sl is}\;X-Y, Y1\;\mbox{\sl is}\;Y+1, q(Z,Y1).
\end{eqnarray*}
We would like to study termination of this program with respect to
$\{q(z_1,z_2)\mid z_1, z_2\;\mbox{\rm are}$ $\mbox{\rm integers}\}$.
Our algorithm infers the following termination condition:
$q_1\leq q_2 \vee (q_1 > q_2 \wedge q_2\geq 0)$. This is a correct
termination condition, but it is not optimal
as $q(z_1,z_2)$ terminates, in fact, for all values of $z_1$ and $z_2$,
i.e., the optimal termination condition is {\em true}.
$\hfill\Box$\end{example}

\section{Experimental evaluation}
The algorithm presented in Figure~\ref{algo} was integrated in the 
system implementing the constraint-based approach 
of~\cite{Decorte:DeSchreye:Vandecasteele}. As a preliminary step
of our analysis, given a program and a set of atomic queries, the 
call set has to be computed. To do so, the type inference technique 
of Janssens and Bruynooghe~\cite{Janssens:Bruynooghe} was used. We opted for a
very simple type inference technique that provides us only with information 
whether some argument is integer or not. More refined analysis can be used.
For instance, the technique presented in \cite{Janssens:Bruynooghe:Englebert}
would have allowed us to know whether some numerical argument belongs to 
a certain interval. Alternatively, the integer intervals domain fo Cousot
and Cousot~\cite{CousotCousot76-1,CousotCousot77-1} might have been used.

We have tested our system on a number of examples. First, we considered
examples from two textbooks chapters
dedicated to programming with arithmetic, namely, Chapter 8 of Sterling 
and Shapiro~\cite{Sterling:Shapiro} and Chapter 9 of Apt~\cite{Apt:Book}.
These results are summarised in Tables~\ref{table:results:SS} and
\ref{table:results:Apt}, respectively. We can prove termination of all
the examples presented for all possible values of the integer arguments,
that is, the termination condition inferred is {\sl true}. 
Next, we've collected a number of programs from different sources. 
Table~\ref{table:results:others} presents timings and results for these 
programs. Again, termination of almost all programs can be shown
for all possible values of the integer arguments. We believe that the
reason for this is that most textbooks authors prefer to write programs 
ensuring termination. 
Finally, Table~\ref{table:results:inference}
demonstrates some of the termination conditions inferred by our system.
We can summarise our results
by saying that the system turned out to be powerful enough to analyse
correctly a broad spectrum of programs, while the time spent on the analysis
never exceeded 0.20 seconds. In fact, for 90\% of the programs results were 
obtained in 0.10 seconds or less. 

The core part of the implementation was done in SICStus 
Prolog~\cite{SICStus:Manual}, type inference of Janssens and 
Bruynooghe~\cite{Janssens:Bruynooghe} was implemented in 
MasterProLog~\cite{MasterProLog}.
Tests were performed on SUN SPARC Ultra-60, model 2360. The 
SPECint\_95 and SPECfp\_95 ratings for this machine are 16.10 and
29.50, respectively. 

\begin{table}
\caption{Examples of Sterling and Shapiro}
\begin{center}
\begin{tabular}{lll}
\hline {\bf Ref}    & {\bf Queries} & {\bf Time} \\
\hline 
8.1    & greatest\_common\_divisor($i$, $i$, $v$) & 0.03\\
8.2    & factorial($i$, $v$)    & 0.02\\
8.3    & factorial($i$, $v$)    & 0.03\\
8.4    & factorial($i$, $v$)    & 0.03 \\
8.5    & between($i$, $i$, $v$) & 0.03 \\
8.6a   & sumlist($li$, $v$)     & 0.00 \\
8.6b   & sumlist($li$, $v$)     & 0.00 \\
8.7a   & inner\_product($li$, $li$, $v$) & 0.00\\
8.7b   & inner\_product($li$, $li$, $v$) & 0.01\\
8.8    & area($lp$, $v$)      &  0.03 \\
8.9    & maxlist($li$, $v$)   &  0.02 \\
8.10   & length($v$, $li$)    &  0.01 \\
8.11   & length($li$, $v$)    &  0.01 \\
8.12   & range($i$, $i$, $v$) &  0.03 \\
\hline
\end{tabular}
\end{center}
\label{table:results:SS}
\end{table}

\begin{table}[htb]
\caption{Examples of Apt}
\begin{center}
\begin{tabular}{lll}
\hline {\bf Name}  & {\bf Queries} & {\bf Time} \\
\hline between    & between($i$, $i$, $v$) & 0.02\\
 delete     & delete ($i$, $i$, $v$) & 0.04\\
 factorial  & fact($i$, $v$)     & 0.01\\
 in\_tree   & in\_tree($i$, $t$) & 0.01\\
 insert  & insert($i$, $t$, $v$) & 0.01\\
 length1 & length($li$, $v$)     & 0.00\\
 maximum & maximum($li$, $v$)    & 0.00\\
 ordered & ordered($li$)         & 0.01\\
 quicksort & qs($li$, $v$)       & 0.10\\
 quicksort\_acc & qs\_acc($li$, $v$)  & 0.10\\
 quicksort\_dl & qs\_dl($li$, $v$)   & 0.13\\
 search\_tree & is\_search\_tree($t$) & 0.06\\
 tree\_minimum & minimum($t$, $v$)    & 0.01\\
\hline
\end{tabular}
\end{center}
\label{table:results:Apt}
\end{table}

\begin{table}[htb]
\caption{Various examples}
\begin{center}
\begin{tabular}{lllll}
\hline {\bf Name} & {\bf Ref} & {\bf Queries} & {\bf Time} & {\bf T} \\ 
\hline dldf &\cite{Bratko} & depthfirst2($c$, $v$, $i$) & 0.03 & T\\ 
 exp & \cite{Coelho:Cotta}& exp($i$, $i$, $v$) & 0.07 & N+ \\ 
 fib &\cite{Maria:Benchmarks} & fib($i$, $v$) & 0.16 & T \\ 
 fib &\cite{OKeefe}& fib($i$, $v$) & 0.05 & T$^{*}$ \\ 
 forwardfib &\cite{Bratko} & fib3($i$, $v$) & 0.02 & T \\ 
 money &\cite{Maria:Benchmarks} & money($v$, $v$, $v$, $v$, & 0.20 & T \\ 
       &                        & \hspace{0.5cm} $v$, $v$, $v$, $v$) & & \\
 oscillate & Example \ref{example:osc} & p($i$) & 0.07 & T \\ 
 p32 &\cite{Hett}& gcd($i$, $i$, $v$)   & 0.03 & T \\ 
 p33 &\cite{Hett}& coprime($i$, $i$)   & 0.05  & T \\ 
 p34 &\cite{Hett}& totient\_phi($i$, $v$)   & 0.14 & T \\ 
 primes &\cite{Clocksin:Mellish} & primes($i$, $v$)   & 0.08 & T \\ 
 pythag &\cite{Clocksin:Mellish} & pythag($v$, $v$, $v$) & 0.05 & N+ \\
 r &\cite{Dershowitz:Lindenstrauss:Sagiv:Serebrenik}& r($i$, $v$) & 0.01 & T\\
 triangle &\cite{McDonald:Yazdani}& triangle($i$, $v$) & 0.03 & N+  \\ 
\hline
\end{tabular}
\end{center}
\label{table:results:others}
\end{table}

\begin{table}[htb]
\caption{Examples of inferring termination conditions}
\begin{center}
\begin{tabular}{lllll}
\hline {\bf Name} & {\bf Ref} & {\bf Queries} & {\bf Time} & {\bf Condition}\\
\hline 
q & Example~\ref{example:gcd:simplified:2} & q($i$, $i$) & 0.04 & $q_1 \leq q_2 \vee (q_1 > q_2 \wedge q_2 > 0)$ \\
q & Example~\ref{example:gcd:simplified:2a} & q($i$, $i$) & 0.05 & $q_1 \leq q_2 \vee (q_1 > q_2 \wedge q_2 \geq 0)$ \\
gcd &\cite{Bratko} & \mbox{\sl gcd}($i$,$i$,$v$) & 0.10 & $q_1 = q_2 \vee (q_1 > q_2 \wedge q_2 \geq 1)$\\ 
\hline
\end{tabular}
\end{center}
\label{table:results:inference}
\end{table}

In Tables~\ref{table:results:SS}--\ref{table:results:inference} the following abbreviations are used:
\begin{itemize}
\item Ref: reference to the program; 
\item Name: name of the program;
\item Queries: single predicate set of atomic queries of interest, where 
the arguments are denoted
\begin{itemize}
\item $c$, if the argument is a character;
\item $i$, if the argument is an integer;
\item $li$, if the argument is a list of integers;
\item $lp$, if the argument is a list of pairs of integers;
\item $t$, if the argument is a binary tree, containing integers;
\item $v$, if the argument is a variable;
\end{itemize} 
\item Time: time (in seconds) needed to analyse the example;
\item T: termination behaviour, see further.
\end{itemize}

One of the less expected results was finding non-terminating examples 
in Prolog textbooks. The first one, due to Coelho and 
Cotta~\cite{Coelho:Cotta}, should compute an $n$th power of a number.
\[\begin{array}{l}
 \mbox{\sl exp}(X,0,1).\\
 \mbox{\sl exp}(X,Y,Z) \leftarrow \mbox{\sl even}(Y), R\;\mbox{\sl is}\;Y/2, P\;\mbox{\sl is}\;X*X, \mbox{\sl exp}(P,R,Z).\\
 \mbox{\sl exp}(X,Y,Z) \leftarrow T\;\mbox{\sl is}\;Y-1, \mbox{\sl exp}(X,T,Z1), Z\;\mbox{\sl is}\;Z1*X.\\
 \\
 \mbox{\sl even}(Y) \leftarrow R\;\mbox{\sl is}\;Y\;\mbox{\sl mod}\;2, R=0.
\end{array}\]
The termination condition inferred by our system is {\sl false} and indeed,
this is the only termination condition possible, since for any goal $G$ the
 LD-tree of this program and $G$ is infinite. This fact is denoted in 
Table~\ref{table:results:others} as N+. Similarly, the fact that for
any goal $G$ the LD-tree of the program and $G$ is finite and our system is
powerful enough to discover this is denoted as T. 

McDonald and Yazdani suggest  the 
following exercise in their book~\cite{McDonald:Yazdani}: 
write a predicate $\mbox{\sl triangle}$ which finds 
the number of balls in a triangle of base $N$. For example, for $N = 4$
the number of balls is $4+3+2+1 = 10$. The next program is the solution
provided by the authors:
\[\begin{array}{l}
 \mbox{\sl triangle}(1,1). \\
 \mbox{\sl triangle}(N,S) \leftarrow M\;\mbox{\sl is}\;N-1, 
      \mbox{\sl triangle}(M,R), S\;\mbox{\sl is}\;M+R.
\end{array}\]
Once more, the termination condition inferred by our system is {\sl false},
and it is the only possible one.

\eat{
One of the most interesting applications of numerical computations we
have found is a depth-first search algorithm of Bratko~\cite{Bratko}. Since
in general depth-first search process may be non-terminating Bratko suggested
to limit the search depth. The following program implements the algorithm.
\[\begin{array}{l}
 \mbox{\sl depthfirst2}(\mbox{\sl Node}, [\mbox{\sl Node}], \_) \leftarrow \mbox{\sl goal}(\mbox{\sl Node}).\\
 \mbox{\sl depthfirst2}(\mbox{\sl Node}, [\mbox{\sl Node}|\mbox{\sl Sol}], \mbox{\sl Maxdepth}) \leftarrow \mbox{\sl Maxdepth} > 0, \\
 \hspace{2.0cm} \mbox{\sl s}(\mbox{\sl Node}, \mbox{\sl Node1}),
             \mbox{\sl Max1}\;\;\mbox{\sl is}\;\;\mbox{\sl Maxdepth} - 1,\\
 \hspace{2.0cm} \mbox{\sl depthfirst2}(\mbox{\sl Node1}, \mbox{\sl Sol},\mbox{\sl  Max1}).
\end{array}\]
Edges of the search tree are represented by atoms $\mbox{\sl s}(\mbox{\sl Node}, \mbox{\sl Node1})$. In order to study termination of this algorithm we 
augmented it with a search tree presented at Figure 11.9~\cite{Bratko}.
}

O'Keefe~\cite{OKeefe} suggested a more efficient way of calculating Fibonacci
numbers performing $O(n)$ work each time it is called, unlike the version 
of~\cite{Maria:Benchmarks} that performs
an exponential amount of work each time.
\[\begin{array}{l}
 \mbox{\sl fib}(1, X)\leftarrow !, X = 1. \\
 \mbox{\sl fib}(2, X)\leftarrow !, X = 1. \\
 \mbox{\sl fib}(N, X)\leftarrow N > 2,  \mbox{\sl fib}(2, N, 1, 1, X).\\
 \\
 \mbox{\sl fib}(N, N, X2, \_, X) \leftarrow !, X = X2. \\
 \mbox{\sl fib}(N_0, N, X2, X1, X) \leftarrow N1\;\mbox{\sl is}\;N_0+1, \\
 \hspace{1.0cm} X3\;\mbox{\sl is}\;X2+X1, \mbox{\sl fib}(N_1, N, X3, X2, X).
\end{array}\]
Termination of goals of the form $\mbox{\sl fib}($i$, $v$)$ with respect
to this example depends on the cut in the first clause of $\mbox{\sl fib}/5$.
If it is removed and if we add at the beginning of the second clause 
$N_0 \neq N$ termination can be proved. This fact is 
denoted T$^{*}$ in Table~\ref{table:results:others}. Note that this 
replacement does not affect complexity of the computation.
Observe also that a more declarative way to write the program might
be to add $N_0 < N$ instead of $N_0 \neq N$. However, while the latter
condition can be inferred automatically from the program, it is not clear 
whether this is also the case for the former one.

\section{Further extensions}
In this section we discuss possible extensions of the algorithm presented 
in Section~\ref{section:generating}. 
First of all, we
discuss integrating termination analysis of numerical and symbolic 
computations, and then show how our results can be used to improve 
existing termination analyses of symbolic computations, such as~\cite{Mesnard,%
Codish:Taboch}. 

\eat{
\subsection{Once more about the inference of adornments}
The set of adornments ${\cal A}_p$, inferred in 
Subsection~\ref{subsection:inferring} may sometimes be too weak
for inferring precise termination conditions. 
\begin{example}
\label{example:permuted}
Consider the following program:
\[\begin{array}{l}
 p(X,Y) \leftarrow X < 0, Y1\;\mbox{\sl is}\;Y+1, X1\;\mbox{\sl is}\;X-1, p(Y1,X1).
\end{array}\]
The maximal integer prefix of the rule above is $X < 0$, thus,
${\cal C}_p = \{p_1 < 0\}$ and
${\cal A}_p = \{p_1 < 0, p_1 \geq 0\}$. The only termination condition
found is $p_1 \geq 0$, while the precise termination
condition is $p_1 \geq 0 \vee (p_1 < 0 \wedge p_2\geq -1)$.
$\hfill\Box$\end{example}

The problem occurred due to the fact that ${\cal A}_p$ restricts 
only {\em some} subset of integer argument positions, while 
for the termination proof information on integer arguments outside of 
this subset may be needed. Thus, we need to infer information
on some variables, given some information on some other variables.

This is to be done by means of interargument relations. 
For the built-in predicates interargument relations follow
from the predefined semantics, (e.g.
$\{(m,n)\mid \mbox{\rm numerical values of $m$ and $n$ are equal}\}$ 
for $\mbox{\sl is}/2$) and for the user-defined predicates the
relations can be inferred similarly to the existing 
techniques~\cite{Decorte:DeSchreye:Vandecasteele}. 

\begin{definition}
\label{def:extension}
Let $P$ be a program, let $p$ be a predicate in $P$, let
$C_q$ be a set of symbolic conditions over the integer 
argument positions of $q$, and $C = \cup_{q\in P} C_q$.
A symbolic condition $c$ over the integer argument positions 
of $p$ is called an 
{\em extension of $C$} if there exists $r\in P$, defining $p$,
such that
some integer argument position denominator appearing in $c$
does not appear in $C_p$, and
$c$ is implied by some $c_q\in C_q$ for the 
recursive subgoals and some interargument relations for the 
non-recursive ones.
\end{definition}
Let $C$ be a set of symbolic conditions over the integer 
argument positions of $p$ and let $\varphi(C)$ be
$C \cup \{c\mid c\;\mbox{\rm is an extension of $C$}\}$. 
Define the set of adornments for $p$ as
$\{c'_1\wedge\ldots\wedge c'_n\mid c'_i\in \varphi^{*}({\cal C}_p)\;\mbox{\rm or}\;\neg c'_i\in \varphi^{*}({\cal C}_p)\}$, where $\varphi^{*}$ is a 
fixpoint of powers of $\varphi$ and ${\cal C}_p$ is defined as in Subsection~\ref{subsection:inferring}.

\begin{example}
Example~\ref{example:permuted}, continued. The only extension of 
${\cal C}_p$ is $p_1 < 0 \wedge p_2 < -1$,
\eat{
.
, since it contains a new integer argument position denominator that does 
not appear in ${\cal C}_p$, i.e., $p_2$, and since it is implied
by taking $p_1 < 0$ for the recursive subgoal $p(Y1,X)$ and
interargument relations for $X>0$ and $Y1\;\mbox{\sl is}\;Y+1$.
This is the only extension of ${\cal C}_p$}
i.e.,
$\varphi({\cal C}_p) = \{p_1 < 0, p_1 < 0 \wedge p_2 < -1\}$.
Thus, 
$\varphi^{*}({\cal C}_p) = \varphi({\cal C}_p)$ and 
${\cal A}_p = \{p_1 < 0 \wedge p_2 < -1, (p_1 < 0 \wedge p_2 \geq -1) \vee p_1\geq 0\}$. 
$\hfill\Box$\end{example}

An alternative approach to propagating such information 
was suggested in~\cite{Dershowitz:Lindenstrauss:Sagiv:Serebrenik}. 
To capture interaction between the variables
a graph was constructed with integer argument positions as vertices and 
a ``can influence''-relation as edges. 
It allows one to propagate the
existing adornments but not to infer the new ones and thus, is less 
precise than our approach.
In the example above
this will allow to infer $\{p_1 < 0 \wedge p_2 < 0, p_1 < 0 \wedge p_2 \geq 0, 
p_1 \geq 0 \wedge p_2 < 0, p_1 \geq 0 \wedge p_2 \geq 0\}$, being less 
precise than the one inferred above. 

It should also be noted that for any inference technique there is an example
where it breaks down. However, we believe that for the majority of practical
examples the methods suggested are powerful enough to infer precise conditions.
}

\eat{
\subsection{Towards general term orderings}
Instead of basing our research on the previous results on acceptability
we could use similar results on term-acceptability~\cite{Serebrenik:DeSchreye:LOPSTR2000}, i.e., acceptability with respect to well-founded term 
orderings. 
This approach allowed to analyse examples such as {\sl derivative}~\cite{DM79:cacm}, {\sl distributive law}~\cite{Dershowitz:Hoot}, {\sl boolean ring}~\cite{Hsiang} and others.

One of the general term orderings that can be extremely useful is
the lexicographic ordering on the set of strings. 
In our case instead of strings vectors of primitive
level mappings are used. 
\begin{example}
This program computes a function, similar
to the Ackermann's function.
\[\begin{array}{l}
 \mbox{\sl sack}(10,N,N1) \leftarrow N1\;\mbox{\sl is}\;N+1.\\
 \mbox{\sl sack}(M1,10,V)\leftarrow M1 < 10, M\;\mbox{\sl is}\;M1+1, \mbox{\sl sack}(M,11,V).\\
 \mbox{\sl sack}(M1,N1,V)\leftarrow M1 < 10, N1 < 10, N\;\mbox{\sl is}\;N1+1, \mbox{\sl sack}(M1,N,V1),\\
 \hspace{2.9cm} M\;\mbox{\rm is}\;M1+1, \mbox{\sl sack}(M,V1,V).
\end{array}\]
The transformation starts as usual and the following adornments are inferred
$\{((\mbox{\sl sack}_1 < 10 \wedge \mbox{\sl sack}_2 > 10) \vee \mbox{\sl sack}_1 > 10),(\mbox{\sl sack}_1 < 10 \wedge \mbox{\sl sack}_2 < 10),
(\mbox{\sl sack}_1 < 10 \wedge \mbox{\sl sack}_2 = 10),\mbox{\sl sack}_1 \geq 10\}$. From the clauses obtained
after adorning two are recursive. For the sake of simplicity we discuss
only one of them, the second one is analysed analogously.
\[\begin{array}{l}
 \mbox{\sl sack}^{\mbox{\sl sack}_1 < 10 \wedge \mbox{\sl sack}_2 < 10}(M1,N1,V)\leftarrow M1 < 10, N1 < 10, N\;\mbox{\rm is}\;N1+1, \\ 
 \hspace{1.5cm} \mbox{\sl sack}^{\mbox{\sl sack}_1 < 10 \wedge \mbox{\sl sack}_2 = 10}(M1,N,V1), M\;\mbox{\rm is}\;M1+1, \mbox{\sl sack}^{\mbox{\sl sack}_1 < 10 \wedge \mbox{\sl sack}_2 < 10}(M,V1,V).
\end{array}\]
Let $\succ$ be defined as
$\mbox{\sl sack}^{\mbox{\sl sack}_1 < 10 \wedge \mbox{\sl sack}_2 < 10}(m_1,n_1,v_1)\;\succ\;
\mbox{\sl sack}^{\mbox{\sl sack}_1 < 10 \wedge \mbox{\sl sack}_2 < 10}(m_2,n_2,v_2)$, 
if either $10 - m_1 > 10 - m_2$ or $(10 - m_1 = 10 - m_2) \wedge
(10 - n_1 > 10 - n_2)$. 
The ordering is well-founded by the well-foundedness of the naturals.
One can easily see that the following decreases hold, implying termination:
\[\begin{array}{l}
 \mbox{\sl sack}^{\mbox{\sl sack}_1 < 10 \wedge \mbox{\sl sack}_2 < 10}(M1,N1,V) \;\succ\;
   \mbox{\sl sack}^{\mbox{\sl sack}_1 < 10 \wedge \mbox{\sl sack}_2 < 10}(M1,N,V1),\;\mbox{\rm where $N = N1+1$}\\
 \mbox{\sl sack}^{\mbox{\sl sack}_1 < 10 \wedge \mbox{\sl sack}_2 < 10}(M1,N1,V) \;\succ\;
   \mbox{\sl sack}^{\mbox{\sl sack}_1 < 10 \wedge \mbox{\sl sack}_2 < 10}(M,V1,V),\;\mbox{\rm where $M = M1+1$}\;\Box
\end{array}\]
$\hfill\Box$\end{example}
} 

\subsection{Integrating numerical and symbolic computation}
In the real-world programs numerical computations are sometimes
interleaved with symbolic ones, as illustrated by the following example 
collecting leaves of a tree with a variable branching factor, 
being a common data structure in natural language processing~\cite{Pollard:Sag}.
\begin{example}
\label{example:collect}
\begin{eqnarray}
&& \mbox{\sl collect}(X,[X|L],L)\leftarrow \mbox{\sl atomic}(X).\nonumber\\
&& \mbox{\sl collect}(T,L0,L) \leftarrow \mbox{\sl compound}(T), \mbox{\sl functor}(T,\_,A), \label{collect:2}\\
&& \hspace{1.0cm} \mbox{\sl process}(T, 0, A, L0, L).\nonumber \\
&& \mbox{\sl process}(\_,A,A,L,L).\nonumber \\
&& \mbox{\sl process}(T,I,A,L0,L2) \leftarrow
        I < A,
        I1\;\mbox{\sl is}\;I+1,
        \mbox{\sl arg}(I1, T, \mbox{\sl Arg}),\label{collect:4} \\
&& \hspace{1.0cm} \mbox{\sl collect}(\mbox{\sl Arg}, L0,L1), 
        \mbox{\sl process}(T,I1,A,L1,L2).\nonumber
\end{eqnarray}
\eat{ 
\[
\begin{array}{ll}
\mbox{\sl collect}(X,[X|L],L)\leftarrow \mbox{\sl atomic}(X). 
& \mbox{\sl process}(\_,A,A,L,L). 
\\
 \mbox{\sl collect}(T,L0,L) \leftarrow 
& \mbox{\sl process}(T,I,A,L0,L2) \leftarrow 
\\
\hspace{0.5cm}\mbox{\sl compound}(T), \mbox{\sl functor}(T,\_,A),
& \hspace{0.5cm} I < A,I1\;\mbox{\sl is}\;I+1,\mbox{\sl arg}(I1, T, \mbox{\sl Arg}),\\
\hspace{0.5cm} \mbox{\sl process}(T, 0, A, L0, L). 
& \hspace{0.5cm}\mbox{\sl collect}(\mbox{\sl Arg}, L0,L1), 
\\
& \hspace{0.5cm} \mbox{\sl process}(T,I1,A,L1,L2). 
\end{array}
\]
} 
To prove termination
 of $\{\mbox{\sl collect}(\mbox{\rm t}, \mbox{\rm v}, [])\}$,
where $t$ is a tree and $v$ is a  variable, 
three decreases should be shown: between a call to
{\sl collect} and a call to {\sl process} in (\ref{collect:2}),
between a call to {\sl process} and a call to {\sl collect} in 
(\ref{collect:4})
and between two calls to {\sl process} in (\ref{collect:4})
. 
The first
two can be shown only by a symbolic level mapping, the third
one---only by the numerical approach. 
$\hfill\Box$\end{example}

Thus, our goal is to {\sl combine} the existing symbolic approaches with
the numerical one presented so far. One of the possible ways to do so is
to combine two level mappings, $\mid\cdot\mid_1$ and $\mid\cdot\mid_2$
by mapping each atom $A\in B^E_P$ either to a natural number $\mid\!A\mid_1$
or to a pair of natural numbers $(\mid\!A\mid_1,\mid\!A\mid_2)$ and prove 
termination by establishing decreases via orderings on 
$({\cal N} \cup {\cal N}^2)$ as suggested in~\cite{Serebrenik:DeSchreye:LOPSTR2000}.
\begin{example}
Example~\ref{example:collect}, continued. Define $\varphi: B^E_P \rightarrow
({\cal N} \cup {\cal N}^2)$ as follows: $\varphi(\mbox{\sl collect}(t,l0,l)) = \|t\|$, 
$\varphi(\mbox{\sl process}(t,i,a,l0,l))  = (\|t\|, a-i)$
where $\|\cdot\|$ is a term-size norm. The decreases
are satisfied with respect to $>$,
such that $A_1 > A_2$ if and only if $\varphi(A_1) \succ \varphi(A_2)$, where
$\succ$ is defined as: $n \succ m$, if $n >_{\cal N} m$,
$n \succ (n, m)$, if {\sl true}, $(n,m_1) \succ (n,m_2)$, if 
$m_1 >_{\cal N} m_2$ and $(n_1,m) \succ n_2$, if $n_1 >_{\cal N} n_2$
and $>_{\cal N}$ is the usual order on the naturals. 
$\hfill\Box$\end{example}

This integrated approach allows one to analyse correctly examples such as
{\sl ground}, {\sl unify}, {\sl numbervars}~\cite{Sterling:Shapiro} and
Example 6.12 in~\cite{Dershowitz:Lindenstrauss:Sagiv:Serebrenik}. 
\eat{ 
some numerical examples, such as Ackermann's function, 
that cannot be analysed by extending~\cite{Decorte:DeSchreye:Vandecasteele}
due to the limitations of level mappings defined as linear combinations, 
can be analysed by the integrated approach.
} 

\subsection{Termination of symbolic computations---revised}
\label{subsection:symrev}
A number of modern approaches to termination analysis of logic programs~\cite{Codish:Taboch,Mesnard,Mesnard:Payet:Neumerkel} abstract a program to CLP($\mbox{\cal N}$) and then infer termination of the original program from the corresponding property of the abstract one. However, as mentioned in the introduction, techniques used to
prove termination of the numerical program are often restricted to
the identity function as the level-mapping. 

\begin{example}
\label{example:less7:terms}
Consider the following example:
\begin{eqnarray*}
&& p(X) \leftarrow \mbox{\sl append}(X,\_, [\_,\_,\_,\_,,\_,\_,\_]), p([\_|X]).\\
&& \mbox{\sl append}([],L,L).\\
&& \mbox{\sl append}([H|X],Y,[H|Z])\leftarrow \mbox{\sl append}(X,Y,Z).
\end{eqnarray*}
Using the list-length norm, defined as 
\[
\|t\| = \left\{
\begin{array}{ll}
1 + \|t'\| &\mbox{\rm if $t=[h|t']$}\\
0          &\mbox{\rm otherwise}
\end{array}
\right.
\]
the following CLP(${\cal N}$)-abstraction can be computed:
\begin{eqnarray*}
&& p(X) \leftarrow \mbox{\sl append}(X,\_,7), p(1+X).\\
&& \mbox{\sl append}(0,L,L).\\
&& \mbox{\sl append}(1+X,Y,1+Z)\leftarrow \mbox{\sl append}(X,Y,Z).
\end{eqnarray*}
Computing a model for the abstraction of $\mbox{\sl append}$ and
transforming the clause for $p$ as described by Mesnard~\cite{Mesnard}
the following program is obtained:
\begin{eqnarray*}
&& p(X) \leftarrow X\leq 7, p(1+X).
\end{eqnarray*}
Termination of this program cannot be shown by the identity function as a level-mapping. Thus, non-termination will be suspected.
$\hfill\Box$\end{example}

Our approach is able to bridge the gap and provide the correct analysis
of Example~\ref{example:less7:terms}. Since  
our results have been stated for numerical computations
and not for CLP(${\cal N}$) minor changes in the abstraction process are
required. Instead of
replacing a term $t$ in an atom $a$ with the size of $t$, 
a fresh variable $V$ is introduced. Then, we add a goal 
$V\;\mbox{\sl is}\;\mbox{\sl size}(t)$ before $a$ (if $a$ is a body subgoal)
or after $a$ (if it is a head of the clause). Next, we replace $t$
in $a$ by $V$, and proceed with the transformation of~\cite{Mesnard}.

\begin{example}
Example~\ref{example:less7:terms}, continued.
Applying the abstraction technique above with respect to the list-length norm
the following program is obtained:
\begin{eqnarray*}
&& p(X) \leftarrow \mbox{\sl append}(X,\_, 7), X1\;\mbox{\sl is}\;X+1, p(X1).\\
&& \mbox{\sl append}(0,L,L).\\
&& \mbox{\sl append}(X1,Y,Z1)\leftarrow X1\;\mbox{\sl is}\;X+1, Z1\;\mbox{\sl is}\;Z+1, \mbox{\sl append}(X,Y,Z).
\end{eqnarray*}
After computing the models this program is transformed to:
\begin{eqnarray*}
&& p(X) \leftarrow X\leq 7, X1\;\mbox{\sl is}\;X+1, p(X1).
\end{eqnarray*}
Termination of $p(n)$ with respect to this program
can be shown by our approach for any integer number $n$. 
This implies termination of $p(t)$ with respect to the original one
for any list of finite length $t$.
$\hfill\Box$\end{example}

To summarise this discussion, we believe that integrating our technique for proving termination of numerical computations with CLP(${\cal N}$) abstracting methodologies of~\cite{Codish:Taboch,Mesnard,Mesnard:Payet:Neumerkel} will significantly extend the class of logic programs that can be analysed automatically. 

\section{Conclusion}
We have presented an approach to verification of termination for 
logic programs with integer computations. This functionality
is lacking in current available termination analysers for Prolog,
such as cTI~\cite{Mesnard,Mesnard:Neumerkel}, 
TerminWeb~\cite{Codish:Taboch}, and
TermiLog~\cite{Lindenstrauss:Sagiv,Lindenstrauss:Sagiv:Serebrenik}. 
The main contribution of this work is threefold. First, from the theoretical
perspective, our study improves the understanding of termination of numerical 
computations, situates them in the well-known framework of 
acceptability and allows integration with the existing approaches
to termination of symbolic computations. Moreover, our technique can 
be used to strengthen the existing techniques for proving termination of 
symbolic computations.

Second, unlike the majority of works on termination analysis for logic 
programs concerned with termination verification, we go further and do inference, i.e., we infer conditions on integer arguments of the queries that imply termination. To perform the inference task we apply a methodology inspired by the constraints based approach~\cite{Decorte:DeSchreye:Vandecasteele}, i.e., we start by symbolic counterparts of level mappings and interargument relations and infer constraints on the integer arguments from rigid acceptability condition and validity of interargument relations.

Finally, the methodology presented has been integrated in the automatic 
termination analyser of~\cite{Decorte:DeSchreye:Vandecasteele}. It
was shown that our approach is robust enough to prove termination for 
a wide range of numerical examples, including
{\sl gcd} and {\sl mod}~\cite{Dershowitz:Lindenstrauss:Sagiv:Serebrenik}
all examples appearing in Chapter 8 
of~\cite{Sterling:Shapiro} and those appearing in~\cite{Apt:Book}.

Termination of numerical computations was studied by a number of 
authors~\cite{Apt:Book,Apt:Marchiori:Palamidessi,%
Dershowitz:Lindenstrauss:Sagiv:Serebrenik}. 
Apt {\em et al.}~\cite{Apt:Marchiori:Palamidessi} provided a declarative
semantics, so called $\Theta$-semantics, for Prolog programs with first-order
built-in predicates, including arithmetic operations. 
In this framework the property of
strong termination, i.e., finiteness of all LD-trees for all possible goals,
was completely characterised based on an appropriately tuned notion of 
acceptability. This approach provides important theoretical results, but
seems to be difficult to integrate in automatic tools. In~\cite{Apt:Book}
it is claimed that an unchanged acceptability condition can be applied to programs in pure Prolog with arithmetic by defining the level mappings on ground atoms with the arithmetic relation to be zero. This approach ignores
the actual computation, and thus, its applicability is restricted
to programs using arithmetic but whose termination behaviour is not dependent
on their arithmetic part, 
such as {\sl quicksort}. Moreover, there are many programs that terminate only for {\em some\/} queries, such as Example~\ref{example:gcd:simplified:2}. Alternatively, Dershowitz {\em et al.}~\cite{Dershowitz:Lindenstrauss:Sagiv:Serebrenik} extended the query-mapping pairs formalism of~\cite{Lindenstrauss:Sagiv} to deal with numerical computations. However, this approach inherited the disadvantages of~\cite{Lindenstrauss:Sagiv}, such as high computational price, inherent to this approach due to repetitive fixpoint computations.
Moreover, since our approach gains its power from the underlying
framework of~\cite{Decorte:DeSchreye:Vandecasteele}, it
allows one to prove termination of some examples that cannot be analysed correctly by~\cite{Dershowitz:Lindenstrauss:Sagiv:Serebrenik}, similar to {\sl confused delete}~\cite{Decorte:DeSchreye:Vandecasteele}. 

More research has been done on termination analysis for constraint logic
programming~\cite{Colussi:Marchiori:Marchiori,Mesnard,Ruggieri:CLP}. 
Since numerical computations in Prolog should be written
in a way that allows a system to verify their satisfiability we can see 
numerical computations of Prolog as an {\em ideal constraint system}. Thus,
all the results obtained for ideal constraints systems can be applied. 
Unfortunately, the research was either oriented towards theoretical 
characterisations~\cite{Ruggieri:CLP} or restricted to
domains isomorphic to ${\cal N}$~\cite{Mesnard}, such as trees and terms.

In a contrast to the approach 
of~\cite{Dershowitz:Lindenstrauss:Sagiv:Serebrenik} that was
restricted to verifying termination,
we presented a methodology for {\sl inferring} 
termination conditions. It is not clear 
whether and how~\cite{Dershowitz:Lindenstrauss:Sagiv:Serebrenik}
can be extended to infer such conditions.

Numerical computations have been also analysed in the early works on 
termination analysis for imperative languages~\cite{Floyd,Katz:Manna},
considering, as we have already pointed out in Section 1, 
general well-founded domains. However, our approach to automation
differs significantly from these works. Traditionally, the verification
community considered automatic generation of invariants~\cite{Bjorner:Browne:Manna}, while automatic generation of ranking functions (level mappings, in the 
logic programming parlance) just started to emerge~\cite{Colon:Sipma:RankingFunctions,Colon:Sipma:PracticalMethods}. The inherent restriction of the latter
results is that ranking functions have to be linear. Moreover, in order to
perform the analysis of larger programs, such as {\em mergesort}, in a
reasonable amount of time, authors further restricted the ranking functions
to depend on one variable only. Unlike these results, our approach doesn't
suffer from such limitations.

The idea of splitting a predicate into cases was first mentioned by Ullman and 
Van Gelder~\cite{Ullman:van:Gelder}, where existence has been assumed 
of a preprocessor that transformed a set of clauses to the new set, in which
every subgoal unifies with all of the rules for its predicate symbol.
However, neither in this paper, nor in the subsequent one (\cite{Sohn:van:Gelder}) the methodology proposed was presented formally. To the best of our 
knowledge the first formal presentation of splitting in the framework
of termination analysis is due to Lindenstrauss {\em et al.}~\cite{Lindenstrauss:Sagiv:Serebrenik:L}. Unlike these results, a numerical and not a symbolic
domain was considered in the current paper.

The termination condition inferred for 
Example~\ref{example:gcd:simplified:2} is {\em optimal}, i.e.,
it is implied by any other termination condition. Clearly, undecidability
of the termination problem implies that no automatic tool can always guarantee
optimality of the condition inferred. However, verifying if the condition
inferred is optimal seems to be an interesting question, related to 
looping analysis~\cite{Bol:PhD,DeSchreye:Verschaetse:Bruynooghe:Horn,Shen,Shen:Yuan:You,Skordev}. So far, in the context of logic programming, 
optimality of termination conditions inferred has been 
studied by Mesnard {\em et al.}~\cite{Mesnard:Payet:Neumerkel} only 
for symbolic computations.

\eat{
Due to the space restrictions we do not present here 
further extensions of our technique (see~\cite{Serebrenik:DeSchreye:cw308}), 
such as more refined domain inference, including propagating information
from one subset of integer argument positions to another one, and integration 
within the framework of 
acceptability based on general term orderings~\cite{Serebrenik:DeSchreye:LOPSTR2000}. The later extension allows to prove termination  
{\em Ackermann's function}. }
 
\section{Acknowledgement}
Alexander Serebrenik is supported by GOA: ``${LP}^{+}$: a second generation
logic programming language''. We are very grateful to Gerda Janssens and
Vincent Englebert for making their type analysis systems available for us.
Many useful suggestions and helpful comments were proposed to us by anonymous referees---we are much obliged to their careful reading.

\bibliographystyle{acmtrans}
\bibliography{/home/alexande/M.Sc.Thesis/main}

\begin{thebibliography}{}

\bibitem[\protect\citeauthoryear{Apt}{Apt}{1997}]{Apt:Book}
{\sc Apt, K.~R.} 1997.
\newblock {\em From Logic Programming to {P}rolog}.
\newblock Prentice-Hall International Series in Computer Science. Prentice
  Hall.

\bibitem[\protect\citeauthoryear{Apt, Marchiori, and Palamidessi}{Apt
  et~al\mbox{.}}{1994}]{Apt:Marchiori:Palamidessi}
{\sc Apt, K.~R.}, {\sc Marchiori, E.}, {\sc and} {\sc Palamidessi, C.} 1994.
\newblock A declarative approach for first-order built-in's in {P}rolog.
\newblock {\em Applicable Algebra in Engineering, Communication and
  Computation\/}~{\em 5,\/}~3/4, 159--191.

\bibitem[\protect\citeauthoryear{Bj{\o}rner, Browne, and Manna}{Bj{\o}rner
  et~al\mbox{.}}{1997}]{Bjorner:Browne:Manna}
{\sc Bj{\o}rner, N.}, {\sc Browne, A.}, {\sc and} {\sc Manna, Z.} 1997.
\newblock Automatic generation of invariants and intermediate assertions.
\newblock {\em Theoretical Computer Science\/}~{\em 173,\/}~1 (February),
  49--87.

\bibitem[\protect\citeauthoryear{Bol}{Bol}{1991}]{Bol:PhD}
{\sc Bol, R.~N.} 1991.
\newblock Loop checking in logic programming.
\newblock Ph.D. thesis, Universiteit van Amsterdam.

\bibitem[\protect\citeauthoryear{Bossi and Cocco}{Bossi and
  Cocco}{1994}]{Bossi:Cocco}
{\sc Bossi, A.} {\sc and} {\sc Cocco, N.} 1994.
\newblock Preserving universal temination through unfold/fold.
\newblock In {\em Algebraic and Logic Programming}, {G.~Levi} {and}
  {M.~Rodr\'{\i}guez-Artalejo}, Eds. Lecture Notes in Computer Science, vol.
  850. Springer Verlag, 269--286.

\bibitem[\protect\citeauthoryear{Bossi, Cocco, Etalle, and Rossi}{Bossi
  et~al\mbox{.}}{2002}]{Bossi:Cocco:Etalle:Rossi:modular}
{\sc Bossi, A.}, {\sc Cocco, N.}, {\sc Etalle, S.}, {\sc and} {\sc Rossi, S.}
  2002.
\newblock On modular termination proofs of general logic programs.
\newblock {\em Theory and Practice of Logic Programming\/}~{\em 2,\/}~3,
  263--291.

\bibitem[\protect\citeauthoryear{Bossi, Cocco, and Fabris}{Bossi
  et~al\mbox{.}}{1991}]{Bossi:Cocco:Fabris:TAPSOFT}
{\sc Bossi, A.}, {\sc Cocco, N.}, {\sc and} {\sc Fabris, M.} 1991.
\newblock Proving {T}ermination of {L}ogic {P}rograms by {E}xploiting {T}erm
  {P}roperties.
\newblock In {\em Proceedings of {CCPSD-TAPSOFT}'91}. Lecture Notes in Computer
  Science, vol. 494. Springer Verlag, 153--180.

\bibitem[\protect\citeauthoryear{Bossi, Cocco, and Fabris}{Bossi
  et~al\mbox{.}}{1994}]{Bossi:Cocco:Fabris}
{\sc Bossi, A.}, {\sc Cocco, N.}, {\sc and} {\sc Fabris, M.} 1994.
\newblock Norms on terms and their use in proving universal termination of a
  logic program.
\newblock {\em Theoretical Computer Science\/}~{\em 124,\/}~2 (February),
  297--328.

\bibitem[\protect\citeauthoryear{Bratko}{Bratko}{1986}]{Bratko}
{\sc Bratko, I.} 1986.
\newblock {\em {P}rolog programming for {A}rtificial {I}ntelligence}.
\newblock Addison-Wesley.

\bibitem[\protect\citeauthoryear{Bruynooghe, Codish, Genaim, and
  Vanhoof}{Bruynooghe et~al\mbox{.}}{2002}]{Bruynooghe:Codish:Genaim:Vanhoof}
{\sc Bruynooghe, M.}, {\sc Codish, M.}, {\sc Genaim, S.}, {\sc and} {\sc
  Vanhoof, W.} 2002.
\newblock {R}euse of results in termination analysis of typed logic programs.
\newblock In {\em Static Analysis, 9th International Symposium}, {M.~V.
  Hermenegildo} {and} {G.~Puebla}, Eds. Lecture Notes in Computer Science, vol.
  2477. Springer Verlag, 477--492.

\bibitem[\protect\citeauthoryear{Bueno, Garc\'{\i}a de~la Banda, and
  Hermenegildo}{Bueno et~al\mbox{.}}{1994}]{Maria:Benchmarks}
{\sc Bueno, F.}, {\sc Garc\'{\i}a de~la Banda, M.~J.}, {\sc and} {\sc
  Hermenegildo, M.~V.} 1994.
\newblock Effectiveness of global analysis in strict independence-based
  automatic parallelization.
\newblock In {\em Logic Programming, Proceedings of the 1994 International
  Symposium}, {M.~Bruynooghe}, Ed. MIT Press, 320--336.

\bibitem[\protect\citeauthoryear{Clocksin and Mellish}{Clocksin and
  Mellish}{1981}]{Clocksin:Mellish}
{\sc Clocksin, W.~F.} {\sc and} {\sc Mellish, C.~S.} 1981.
\newblock {\em Programming in {P}rolog}.
\newblock {S}pringer {V}erlag.

\bibitem[\protect\citeauthoryear{Codish and Taboch}{Codish and
  Taboch}{1999}]{Codish:Taboch}
{\sc Codish, M.} {\sc and} {\sc Taboch, C.} 1999.
\newblock A semantic basis for termination analysis of logic programs.
\newblock {\em Journal of Logic Programming\/}~{\em 41,\/}~1, 103--123.

\bibitem[\protect\citeauthoryear{Coelho and Cotta}{Coelho and
  Cotta}{1988}]{Coelho:Cotta}
{\sc Coelho, H.} {\sc and} {\sc Cotta, J.~C.} 1988.
\newblock {\em {P}rolog by example}.
\newblock Springer Verlag.

\bibitem[\protect\citeauthoryear{Col{\'o}n and Sipma}{Col{\'o}n and
  Sipma}{2001}]{Colon:Sipma:RankingFunctions}
{\sc Col{\'o}n, M.~A.} {\sc and} {\sc Sipma, H.~B.} 2001.
\newblock Synthesis of linear ranking functions.
\newblock In {\em Tools and Algorithms for the Construction and Analysis of
  Systems, 7th International Conference}, {T.~Margaria} {and} {W.~Yi}, Eds.
  Lecture Notes in Computer Science, vol. 2031. Springer Verlag, 67--81.

\bibitem[\protect\citeauthoryear{Col{\'o}n and Sipma}{Col{\'o}n and
  Sipma}{2002}]{Colon:Sipma:PracticalMethods}
{\sc Col{\'o}n, M.~A.} {\sc and} {\sc Sipma, H.~B.} 2002.
\newblock Practical methods for proving program termination.
\newblock In {\em Computer Aided Verification, 14th International Conference},
  {E.~Brinksma} {and} {K.~Guldstrand~Larsen}, Eds. Lecture Notes in Computer
  Science, vol. 2404. Springer Verlag, 442--454.

\bibitem[\protect\citeauthoryear{Colussi, Marchiori, and Marchiori}{Colussi
  et~al\mbox{.}}{1995}]{Colussi:Marchiori:Marchiori}
{\sc Colussi, L.}, {\sc Marchiori, E.}, {\sc and} {\sc Marchiori, M.} 1995.
\newblock On termination of constraint logic programs.
\newblock In {\em Principles and Practice of Constraint Programming - CP'95,},
  {U.~Montanari} {and} {F.~Rossi}, Eds. Lecture Notes in Computer Science, vol.
  976. Springer Verlag, 431--448.

\bibitem[\protect\citeauthoryear{Cousot and Cousot}{Cousot and
  Cousot}{1976}]{CousotCousot76-1}
{\sc Cousot, P.} {\sc and} {\sc Cousot, R.} 1976.
\newblock Static determination of dynamic properties of programs.
\newblock In {\em Proceedings of the Second International Symposium on
  Programming}. Dunod, Paris, France, 106--130.

\bibitem[\protect\citeauthoryear{Cousot and Cousot}{Cousot and
  Cousot}{1977}]{CousotCousot77-1}
{\sc Cousot, P.} {\sc and} {\sc Cousot, R.} 1977.
\newblock Abstract interpretation: a unified lattice model for static analysis
  of programs by construction or approximation of fixpoints.
\newblock In {\em Conference Record of the Fourth Annual ACM SIGPLAN-SIGACT
  Symposium on Principles of Programming Languages}. ACM Press, New York, NY,
  Los Angeles, California, 238--252.

\bibitem[\protect\citeauthoryear{De~Schreye, Verschaetse, and
  Bruynooghe}{De~Schreye
  et~al\mbox{.}}{1990}]{DeSchreye:Verschaetse:Bruynooghe:Horn}
{\sc De~Schreye, D.}, {\sc Verschaetse, K.}, {\sc and} {\sc Bruynooghe, M.}
  1990.
\newblock A practical technique for detecting non-terminating queries for a
  restricted class of {H}orn clauses, using directed, weighted graphs.
\newblock In {\em Logic Programming, Proceedings of the Seventh International
  Conference}, {D.~H. Warren} {and} {P.~Szeredi}, Eds. MIT Press, 649--663.

\bibitem[\protect\citeauthoryear{De~Schreye, Verschaetse, and
  Bruynooghe}{De~Schreye
  et~al\mbox{.}}{1992}]{DeSchreye:Verschaetse:Bruynooghe}
{\sc De~Schreye, D.}, {\sc Verschaetse, K.}, {\sc and} {\sc Bruynooghe, M.}
  1992.
\newblock A framework for analyzing the termination of definite logic programs
  with respect to call patterns.
\newblock In {\em Proceedings of the International Conference on Fifth
  Generation Computer Systems.}, {I.~Staff}, Ed. IOS Press, 481--488.

\bibitem[\protect\citeauthoryear{Decorte and De~Schreye}{Decorte and
  De~Schreye}{1998}]{Decorte:DeSchreye:98}
{\sc Decorte, S.} {\sc and} {\sc De~Schreye, D.} 1998.
\newblock Termination analysis: some practical properties of the norm and level
  mapping space.
\newblock In {\em Proceedings of the 1998 Joint International Conference and
  Symposium on Logic Programming}, {J.~Jaffar}, Ed. MIT Press, 235--249.

\bibitem[\protect\citeauthoryear{Decorte, De~Schreye, and
  Vandecasteele}{Decorte et~al\mbox{.}}{1999}]{Decorte:DeSchreye:Vandecasteele}
{\sc Decorte, S.}, {\sc De~Schreye, D.}, {\sc and} {\sc Vandecasteele, H.}
  1999.
\newblock Constraint-based termination analysis of logic programs.
\newblock {\em ACM TOPLAS\/}~{\em 21,\/}~6 (November), 1137--1195.

\bibitem[\protect\citeauthoryear{Dershowitz, Lindenstrauss, Sagiv, and
  Serebrenik}{Dershowitz
  et~al\mbox{.}}{2001}]{Dershowitz:Lindenstrauss:Sagiv:Serebrenik}
{\sc Dershowitz, N.}, {\sc Lindenstrauss, N.}, {\sc Sagiv, Y.}, {\sc and} {\sc
  Serebrenik, A.} 2001.
\newblock {A} general framework for automatic termination analysis of logic
  programs.
\newblock {\em Applicable Algebra in Engineering, Communication and
  Computing\/}~{\em 12,\/}~1-2, 117--156.

\bibitem[\protect\citeauthoryear{Floyd}{Floyd}{1967}]{Floyd}
{\sc Floyd, R.~W.} 1967.
\newblock Assigning meanings to programs.
\newblock In {\em Mathematical Aspects of Computer Science}, {J.~Schwartz}, Ed.
  American Mathematical Society, 19--32.
\newblock Proceedings of Symposiumsia in Applied Mathematics; v. 19.

\bibitem[\protect\citeauthoryear{Genaim and Codish}{Genaim and
  Codish}{2001}]{Genaim:Codish}
{\sc Genaim, S.} {\sc and} {\sc Codish, M.} 2001.
\newblock Inferring termination conditions for logic programs using backwards
  analysis.
\newblock In {\em Logic for Programming, Artificial Intelligence, and
  Reasoning, 8th International Conferencerence, Proceedings}, {R.~Nieuwenhuis}
  {and} {A.~Voronkov}, Eds. Lecture Notes in Computer Science, vol. 2250.
  Springer Verlag, 685--694.

\bibitem[\protect\citeauthoryear{Genaim, Codish, Gallagher, and Lagoon}{Genaim
  et~al\mbox{.}}{2002}]{Genaim:Codish:Gallagher:Lagoon}
{\sc Genaim, S.}, {\sc Codish, M.}, {\sc Gallagher, J.}, {\sc and} {\sc Lagoon,
  V.} 2002.
\newblock Combining norms to prove termination.
\newblock In {\em Third International Workshop on Verification, Model Checking
  and Abstract Interpretation}, {A.~Cortesi}, Ed. Lecture Notes in Computer
  Science, vol. 2294. Springer Verlag, 126--138.

\bibitem[\protect\citeauthoryear{Hett}{Hett}{2001}]{Hett}
{\sc Hett, W.} 2001.
\newblock P-99: Ninety-nine {P}rolog problems.
\newblock Available at
  \verb+http://www.hta-bi.bfh.ch/~hew/informatik3/prolog/p-99/+.

\bibitem[\protect\citeauthoryear{Holzbaur}{Holzbaur}{1995}]{CLP:Manual}
{\sc Holzbaur, C.} 1995.
\newblock O{F}{A}{I} {CLP(Q,R)} {M}anual.
\newblock Tech. Rep. TR-95-09, Austrian Research Institute for Artificial
  Intelligence, Vienna.

\bibitem[\protect\citeauthoryear{{ILOG}}{{ILOG}}{2001}]{ILOG}
{\sc {ILOG}}. 2001.
\newblock {\em {ILOG} {S}olver 5.1 User's Manual}.
\newblock {ILOG} s.a. \verb+ http://www.ilog.com+.

\bibitem[\protect\citeauthoryear{{IT M}asters}{{IT
  M}asters}{2000}]{MasterProLog}
{\sc {IT M}asters}. 2000.
\newblock Master{P}ro{L}og {P}rogramming {E}nvironment.
\newblock Available at \verb+http://www.itmasters.com/+.

\bibitem[\protect\citeauthoryear{Janssens and Bruynooghe}{Janssens and
  Bruynooghe}{1992}]{Janssens:Bruynooghe}
{\sc Janssens, G.} {\sc and} {\sc Bruynooghe, M.} 1992.
\newblock {D}eriving descriptions of possible values of program variables by
  means of abstract interpretation.
\newblock {\em Journal of Logic Programming\/}~{\em 13,\/}~2\&3 (July),
  205--258.

\bibitem[\protect\citeauthoryear{Janssens, Bruynooghe, and Englebert}{Janssens
  et~al\mbox{.}}{1994}]{Janssens:Bruynooghe:Englebert}
{\sc Janssens, G.}, {\sc Bruynooghe, M.}, {\sc and} {\sc Englebert, V.} 1994.
\newblock Abstracting numerical values in {CLP(H, N)}.
\newblock In {\em Programming Language Implementation and Logic Programming,
  6th International Symposiumsium, PLILP'94}, {M.~V. Hermenegildo} {and}
  {J.~Penjam}, Eds. Lecture Notes in Computer Science, vol. 844. Springer
  Verlag, 400--414.

\bibitem[\protect\citeauthoryear{Katz and Manna}{Katz and
  Manna}{1975}]{Katz:Manna}
{\sc Katz, S.} {\sc and} {\sc Manna, Z.} 1975.
\newblock A closer look at termination.
\newblock {\em Acta Informatica\/}~{\em 5}, 333--352.

\bibitem[\protect\citeauthoryear{Lindenstrauss and Sagiv}{Lindenstrauss and
  Sagiv}{1997}]{Lindenstrauss:Sagiv}
{\sc Lindenstrauss, N.} {\sc and} {\sc Sagiv, Y.} 1997.
\newblock Automatic termination analysis of logic programs.
\newblock In {\em Proceedings of the Fourteenth International Conference on
  Logic Programming}, {L.~Naish}, Ed. MIT Press, 63--77.

\bibitem[\protect\citeauthoryear{Lindenstrauss, Sagiv, and
  Serebrenik}{Lindenstrauss
  et~al\mbox{.}}{1997}]{Lindenstrauss:Sagiv:Serebrenik}
{\sc Lindenstrauss, N.}, {\sc Sagiv, Y.}, {\sc and} {\sc Serebrenik, A.} 1997.
\newblock {\em TermiLog\/}: A system for checking termination of queries to
  logic programs.
\newblock In {\em Computer Aided Verification, 9th International Conference},
  {O.~Grumberg}, Ed. Lecture Notes in Computer Science, vol. 1254. Springer
  Verlag, 63--77.

\bibitem[\protect\citeauthoryear{Lindenstrauss, Sagiv, and
  Serebrenik}{Lindenstrauss
  et~al\mbox{.}}{1998}]{Lindenstrauss:Sagiv:Serebrenik:L}
{\sc Lindenstrauss, N.}, {\sc Sagiv, Y.}, {\sc and} {\sc Serebrenik, A.} 1998.
\newblock Unfolding the mystery of {\em mergesort\/}.
\newblock In {\em Proceedings of the $7^{th}$ International Workshop on Logic
  Program Synthesis and Transformation}, {N.~Fuchs}, Ed. Lecture Notes in
  Computer Science, vol. 1463. Springer Verlag.

\bibitem[\protect\citeauthoryear{McDonald and Yazdani}{McDonald and
  Yazdani}{1990}]{McDonald:Yazdani}
{\sc McDonald, C.} {\sc and} {\sc Yazdani, M.} 1990.
\newblock {\em {P}rolog programming: a tutorial introduction}.
\newblock Artificial Intelligence Texts. Blackwell Scientific Publications.

\bibitem[\protect\citeauthoryear{Mesnard}{Mesnard}{1996}]{Mesnard}
{\sc Mesnard, F.} 1996.
\newblock Inferring left-terminating classes of queries for constraint logic
  programs.
\newblock In {\em Proceedings of the 1996 {J}oint {I}nternational {C}onference
  and {S}yposium on {L}ogic {P}rogramming}, {M.~Maher}, Ed. The {MIT} {P}ress,
  Cambridge, {MA}, {USA}, 7--21.

\bibitem[\protect\citeauthoryear{Mesnard and Neumerkel}{Mesnard and
  Neumerkel}{2001}]{Mesnard:Neumerkel}
{\sc Mesnard, F.} {\sc and} {\sc Neumerkel, U.} 2001.
\newblock Applying static analysis techniques for inferring termination
  conditions of logic programs.
\newblock In {\em Static Analysis, 8th International Symposium, SAS 2001},
  {P.~Cousot}, Ed. Lecture Notes in Computer Science, vol. 2126. Springer
  Verlag, 93--110.

\bibitem[\protect\citeauthoryear{Mesnard, Payet, and Neumerkel}{Mesnard
  et~al\mbox{.}}{2002}]{Mesnard:Payet:Neumerkel}
{\sc Mesnard, F.}, {\sc Payet, E.}, {\sc and} {\sc Neumerkel, U.} 2002.
\newblock Dectecting optimal termination conditions of logic programs.
\newblock In {\em Static Analysis, 8th International Symposium, SAS 2002},
  {P.~Cousot}, Ed. Lecture Notes in Computer Science, vol. 2477. Springer
  Verlag, 509--527.

\bibitem[\protect\citeauthoryear{Mesnard and Ruggieri}{Mesnard and
  Ruggieri}{2003}]{Mesnard:Ruggieri}
{\sc Mesnard, F.} {\sc and} {\sc Ruggieri, S.} 2003.
\newblock On proving left termination of constraint logic programs.
\newblock {\em {ACM} Transaction on Computational Logic\/}~{\em 4,\/}~2,
  207--259.

\bibitem[\protect\citeauthoryear{Ohlebusch}{Ohlebusch}{2001}]{Ohlebusch}
{\sc Ohlebusch, E.} 2001.
\newblock Automatic termination proofs of logic programs via rewrite systems.
\newblock {\em Applicable Algebra in Engineering, Communication and
  Computing\/}~{\em 12,\/}~1-2, 73--116.

\bibitem[\protect\citeauthoryear{O'Keefe}{O'Keefe}{1990}]{OKeefe}
{\sc O'Keefe, R.~A.} 1990.
\newblock {\em The {C}raft of {P}rolog}.
\newblock {MIT} {P}ress, Cambridge, {MA}, {USA}.

\bibitem[\protect\citeauthoryear{Pl{\"u}mer}{Pl{\"u}mer}{1990}]{Plumer:ICLP}
{\sc Pl{\"u}mer, L.} 1990.
\newblock Termination proofs for logic programs based on predicate
  inequalities.
\newblock In {\em Proceedings of ICLP'90}. MIT Press, 634--648.

\bibitem[\protect\citeauthoryear{Pl{\"u}mer}{Pl{\"u}mer}{1991}]{Plumer}
{\sc Pl{\"u}mer, L.} 1991.
\newblock Automatic termination proofs for {P}rolog programs operating on
  nonground terms.
\newblock In {\em International Logic Programming Symposium}. MIT Press.

\bibitem[\protect\citeauthoryear{Pollard and Sag}{Pollard and
  Sag}{1994}]{Pollard:Sag}
{\sc Pollard, C.} {\sc and} {\sc Sag, I.~A.} 1994.
\newblock {\em Head-driven Phrase Structure Grammar}.
\newblock The University of Chicago Press.

\bibitem[\protect\citeauthoryear{Ruggieri}{Ruggieri}{1997}]{Ruggieri:CLP}
{\sc Ruggieri, S.} 1997.
\newblock Termination of constraint logic programs.
\newblock In {\em Automata, Languages and Programming, 24th International
  Colloquium, ICALP'97}, {P.~Degano}, {R.~Gorrieri}, {and}
  {A.~Marchetti-Spaccamela}, Eds. Lecture Notes in Computer Science, vol. 1256.
  Springer Verlag, 838--848.

\bibitem[\protect\citeauthoryear{Serebrenik and De~Schreye}{Serebrenik and
  De~Schreye}{2001}]{Serebrenik:DeSchreye:LOPSTR2000}
{\sc Serebrenik, A.} {\sc and} {\sc De~Schreye, D.} 2001.
\newblock {N}on-transformational termination analysis of logic programs, based
  on general term-orderings.
\newblock In {\em Logic Based Program Synthesis and Transformation 10th
  International Workshop, Selected Papers}, {K.-K. Lau}, Ed. Lecture Notes in
  Computer Science, vol. 2042. Springer Verlag, 69--85.

\bibitem[\protect\citeauthoryear{Serebrenik and De~Schreye}{Serebrenik and
  De~Schreye}{2002}]{Serebrenik:DeSchreye:real}
{\sc Serebrenik, A.} {\sc and} {\sc De~Schreye, D.} 2002.
\newblock On termination of logic programs with floating point computations.
\newblock In {\em 9th International Static Analysis Symposium}, {M.~V.
  Hermenegildo} {and} {G.~Puebla}, Eds. Lecture Notes in Computer Science, vol.
  2477. Springer Verlag, 151--164.

\bibitem[\protect\citeauthoryear{Shen}{Shen}{1997}]{Shen}
{\sc Shen, Y.-D.} 1997.
\newblock An extended variant of atoms loop check for positive logic programs.
\newblock {\em New Generation Computing\/}~{\em 15,\/}~2, 187--204.

\bibitem[\protect\citeauthoryear{Shen, Yuan, and You}{Shen
  et~al\mbox{.}}{2001}]{Shen:Yuan:You}
{\sc Shen, Y.-D.}, {\sc Yuan, L.-Y.}, {\sc and} {\sc You, J.-H.} 2001.
\newblock Loop checks for logic programs with functions.
\newblock {\em Theoretical Computer Science\/}~{\em 266,\/}~1--2, 441--461.

\bibitem[\protect\citeauthoryear{{SICS}}{{SICS}}{2002}]{SICStus:Manual}
{\sc {SICS}}. 2002.
\newblock {\em {SICS}tus User Manual. Version 3.10.0}.
\newblock Swedish Institute of Computer Science.

\bibitem[\protect\citeauthoryear{Skordev}{Skordev}{1997}]{Skordev}
{\sc Skordev, D.} 1997.
\newblock An abstract approach to some loop detection problems.
\newblock {\em Fundamenta Informaticae\/}~{\em 31,\/}~2, 195--212.

\bibitem[\protect\citeauthoryear{Sohn and Van~Gelder}{Sohn and
  Van~Gelder}{1991}]{Sohn:van:Gelder}
{\sc Sohn, K.} {\sc and} {\sc Van~Gelder, A.} 1991.
\newblock Termination detection in logic programs using argument sizes.
\newblock In {\em Proceedings of the Tenth ACM SIGACT-SIGART-SIGMOD Symposium
  on Principles of Database Systems}. {ACM} {P}ress, 216--226.

\bibitem[\protect\citeauthoryear{Sterling and Shapiro}{Sterling and
  Shapiro}{1994}]{Sterling:Shapiro}
{\sc Sterling, L.} {\sc and} {\sc Shapiro, E.} 1994.
\newblock {\em The {A}rt of {P}rolog}.
\newblock The {MIT} Press, Cambridge, {MA}, {USA}.

\bibitem[\protect\citeauthoryear{Ullman and Van~Gelder}{Ullman and
  Van~Gelder}{1988}]{Ullman:van:Gelder}
{\sc Ullman, J.~D.} {\sc and} {\sc Van~Gelder, A.} 1988.
\newblock Efficient tests for top-down termination of logical rules.
\newblock {\em Journal of the {ACM}\/}~{\em 35,\/}~2 (April), 345--373.

\bibitem[\protect\citeauthoryear{Verbaeten, Sagonas, and De~Schreye}{Verbaeten
  et~al\mbox{.}}{2001}]{Verbaeten:Sagonas:DeSchreye}
{\sc Verbaeten, S.}, {\sc Sagonas, K.}, {\sc and} {\sc De~Schreye, D.} 2001.
\newblock {T}ermination proofs for logic programs with tabling.
\newblock {\em ACM Transactions on Computational Logic\/}~{\em 2,\/}~1, 57--92.

\bibitem[\protect\citeauthoryear{Verschaetse and De~Schreye}{Verschaetse and
  De~Schreye}{1991}]{Verschaetse:DeSchreye}
{\sc Verschaetse, K.} {\sc and} {\sc De~Schreye, D.} 1991.
\newblock Deriving termination proofs for logic programs, using abstract
  procedures.
\newblock In {\em Logic Programming, Proceedings of the Eigth International
  Conference}, {K.~Furukawa}, Ed. MIT Press, 301--315.

\bibitem[\protect\citeauthoryear{Winsborough}{Winsborough}{1992}]{Winsborough}
{\sc Winsborough, W.} 1992.
\newblock Multiple specialization using minimal-function graph semantics.
\newblock {\em Journal of Logic Programming\/}~{\em 13,\/}~2/3, 259--290.

\end{thebibliography}

\end{document}